\documentclass[12pt]{article}
\usepackage{amsmath}
\usepackage{graphicx,psfrag,epsf}
\usepackage{enumerate}
\usepackage{url} 
\usepackage{float}
\usepackage[table,xcdraw,dvipsnames]{xcolor}

\usepackage{cite}  

\usepackage{multirow}
\usepackage{lipsum}
\usepackage{amsfonts}
\usepackage{cite}      
                  
\usepackage{graphicx}   

\usepackage{amsthm}
\RequirePackage[colorlinks,citecolor=blue,urlcolor=blue]{hyperref}
\usepackage{amssymb,calc}
\usepackage{thmtools}
\usepackage{mathrsfs}
\usepackage{mathtools}
\usepackage{bm}
\usepackage{bbm}
\usepackage{caption}
\usepackage{subcaption}
\usepackage{textgreek}

\usepackage{enumerate}
\usepackage{rotating}
\RequirePackage{cleveref}
\usepackage{hyphenat}
\usepackage[round]{natbib}
\usepackage{caption,subcaption}
\usepackage{tikz}
\usepackage{pgfplots}

\usepackage[T1]{fontenc}
\usepackage{textcomp} 
\usepackage{lmodern}

\usepackage{algorithm,algorithmic}
\usepackage[algo2e]{algorithm2e}

\makeatletter
\renewcommand{\ALG@name}{Table}
\makeatother

\pgfplotsset{width=10cm}

\tikzset{declare function={gamma(\x)=sqrt(2*pi)*\x^(\x-0.5)*exp(-\x)*exp(1/(12*\x));}}
\tikzset{declare function={tpdf(\x,\nu)=gamma(0.5*(\nu+1))/(sqrt(pi*\nu)*gamma(\nu/2))*(1+\x^2/\nu)^(-(\nu+1)/2);}}
\tikzset{declare function={invgampdf(\x,\a,\b)=(\b/\x)^\a/\x/gamma(\a)*exp(-\b/\x);}}

\newcommand{\nhphantom}[1]{\ifmmode\settowidth{\dimen0}{$#1$}\else\settowidth{\dimen0}{#1}\fi\hspace*{-\dimen0}}

\usetikzlibrary{patterns}
\tikzset{
	hatch distance/.store in=\hatchdistance,
	hatch distance=5pt,
	hatch thickness/.store in=\hatchthickness,
	hatch thickness=0.5pt,
}

\newcommand{\wh}[1]{\widehat{#1}}

\definecolor{pink}{rgb}{0.9, 0.17, 0.31}

\newcommand\w{\bm w}
\newcommand\x{\bm x}
\def\C {\,|\:}

\newcommand\F{\mathcal F}
\newcommand\bw{\bm w}

\newcommand\E{\mathbb E}

\renewcommand\P{\mathbb P}

\newcommand\Xn{X^{(n)}}
\newcommand\R{\mathbb R}
\renewcommand\b{\bm{\beta}}

\newcommand\iid{\text{iid}}

\newcommand{\wt}[1]{\widetilde{#1}}

\newtheorem{assumption}{Assumption}
\newtheorem{lemma}{Lemma}

\renewcommand{\nhphantom}[1]{\ifmmode\settowidth{\dimen0}{$#1$}\else\settowidth{\dimen0}{#1}\fi\hspace*{-\dimen0}}

\numberwithin{equation}{section}

\newtheorem{thm}{Theorem}[section]

\newtheorem{lem}[thm]{Lemma}

\newtheorem{rem}{Remark}

\crefname{thm}{Theorem}{Theorems}
\crefname{prop}{Proposition}{Propositions}
\crefname{lem}{Lemma}{Lemmas}
\crefname{coro}{Corollary}{Corollaries}
\crefname{add}{Addendum}{Addendums}
\crefname{asm}{Assumption}{Assumptions}
\crefname{alg}{Algorithm}{Algorithms}
\crefname{proc}{Procedure}{Procedures}
\crefname{exe}{Exercise}{Exercises}
\crefname{exa}{Example}{Examples}
\crefname{prob}{Problem}{Problems}
\crefname{section}{Section}{Sections}
\crefname{subsection}{Section}{Sections}
\crefname{appendix}{Appendix}{Appendices}

\allowdisplaybreaks[3]

\begin{document}

\def\spacingset#1{\renewcommand{\baselinestretch}%
{#1}\small\normalsize} \spacingset{1}

	\title{\sf Deep Bootstrap  for Bayesian Inference }
	\author{
		Lizhen Nie\footnote{
			 5th year PhD student at the Department of Statistics of the University of Chicago}  \,\, and    Veronika Ro\v{c}kov\'{a}\footnote{  Associate Professor in Econometrics and Statistics and James S. Kemper Faculty Scholar at the {\sl \small Booth School of Business, University of Chicago}.
			The author gratefully acknowledges the support from the James S. Kemper Faculty Fund at the Booth School of Business and 
			the National Science Foundation (Grant No. NSF DMS-1944740).} 
	} 
	\maketitle

\bigskip
\begin{abstract}
For a Bayesian,  the task to define the likelihood can be as perplexing as the task to define the prior.
We focus on situations when the parameter of interest has been emancipated from the likelihood and is linked to data directly through a loss function.
We survey existing work on both Bayesian parametric inference with Gibbs posteriors as well as  Bayesian non-parametric inference.
We then highlight recent bootstrap computational approaches to approximating loss-driven posteriors. 
In particular,  we focus on implicit  bootstrap distributions defined through an underlying push-forward mapping.
We investigate  iid   samplers from approximate posteriors that pass random bootstrap weights trough a trained  generative network. 
After training the deep-learning mapping, the simulation cost of such iid samplers is negligible.  We compare the performance of these deep bootstrap samplers with exact bootstrap as well as MCMC on several 
examples (including support vector machines or quantile regression). We also provide theoretical insights into bootstrap posteriors by drawing upon connections to model mis-specification.

\end{abstract}

\noindent%

\spacingset{1} 


\section{Introduction}

While Bayesian's obligation to specify a prior has been challenged the most, the obligation to specify the likelihood  is perhaps even more consequential.
Implicit in the  Bayesian paradigm is the assumption that a probabilistic model can be formulated which links parameters with data.
When constructing such a model, however, one has to reconcile   bias implications of model mis-specification (\cite{muller2013risk,grunwald2017inconsistency}). 
Very often,
the primary aim is not  modeling the data but rather estimating a statistic. Examples include M-estimators (\cite{huber2009robust}) or other extremum estimators  such as censored quantile regression (\cite{powell1986censored}), instrumental and robust median regression (\cite{mood1950introduction}), nonlinear IV and GMM (\cite{hansen1982generalized}).  For statistical inference, one may wish to obtain a post-data density summary of the parameter of interest rather than just a point estimate.  
Bayesian  prior-to-posterior  inference can be carried out even when one is reluctant about committing to a particular generative model.

One way how to liberate the inferential parameter $\theta_0$ from the likelihood is by relating it to data through a general loss function. In medicine, for instance, minimal clinically important difference (\cite{syring2017gibbs}) or boundary detection in image analysis (\cite{syring2020robust})
can be formulated as loss minimization  problems. It is then possible to perform coherent Bayesian-style updating of prior beliefs, expressed in the prior $\pi(\theta)$,  through the so-called Gibbs posteriors (\cite{zhang2006epsilon,zhang2006information,bissiri2016general}).  This is a parametric generalization of classical Bayesian inference where the loss function is converted into a pseudo-likelihood function. Another way of  expressing uncertainty about $\theta_0$ is through a prior  $\pi(F)$, as opposed to the prior $\pi(\theta)$, on the unknown  data generating distribution function  $F_0$. Such non-parametric Bayesian inference (\cite{chamberlain2003nonparametric,lyddon2019general}) is based  on the posterior of $F$ rather than $\theta$. We revisit both the non-parametric and parametric Bayesian approach (with Gibbs posteriors) to inference about parameter targets defined through loss functions.

Recalling the optimal information processing interpretation of the Bayes' rule (\cite{zellner1988optimal,knoblauch2019generalized}), this paper surveys various generalizations of Bayesian inference (including variational inference (\cite{jordan1999introduction,wainwright2008graphical}) and Gibbs posteriors (\cite{catoni2004statistical,zhang2006information})) under one unifying hat. 
In particular, we adopt the optimization-centric point of view on Bayes' rule which allows a re-interpretation of Bayesian inference as regularized optimization. 
Any commitment to a Bayesian posterior is a commitment to a particular optimization objective parametrized by the prior, the loss (log-likelihood) function and the class of post-data inferential densities (\cite{knoblauch2019generalized}).  For example, variational Bayes  forces the posterior belief into a specific parametric form, transforming the optimization from an infinite-dimensional into a finite-dimensional one. Gibbs posteriors, on the other hand, force priors and data into an exponentially additive relationship through loss functions.
\cite{alquier2016properties} combined the two by providing a VB computational alternative for Gibbs posteriors.

The repertoire of  sampling methods for computing Gibbs posteriors depends  on the availability of closed-form conditionals or computational resources. For example, classical MCMC sampling (using e.g. Metropolis-style samplers) may   incur  large computational costs (\cite{quiroz2018speeding,johndrow2020scalable}).  As an alternative, this work investigates the
 recently proposed generative bootstrap sampler (\cite{shin2020scalable}) in the context of Bayesian simulation. This generative sampler is trained by  learning a deterministic (deep learning) mapping between bootstrap weights and parameters to perform iid sampling. The iid aspect is particularly appealing because, after the mapping has been trained, the simulation cost is negligible compared to sequential samplers.  We tailor the strategy of \cite{shin2020scalable} to the context of Bayesian simulation from approximate (1) Gibbs posteriors for parametric Bayesian inference, and (2)  non-parametric Bayesian posteriors.
The goal is to learn an implicit distribution  prescribed by a deterministic mapping that
filters out bootstrap weights to produce samples from an approximate posterior. 
Implicit distributions have been loosely defined  as distributions  whose likelihoods are unavailable but which can be sampled from (\cite{mohamed2016learning,li2018gradient}).  While implicit distributions have been deployed for Bayesian computation before in the context of variational Bayes (\cite{ruiz2019contrastive}), we explore implicit bootstrap distributions to generate samples from  approximate posteriors. 

Our main purpose is to (1) highlight several recent developments in loss-based Bayesian inference, and to (2) draw attention to generative samplers which are potentially very promising for Bayesian simulation. We investigate their benefits as well as limitations. We start off by describing loss-based Bayesian inference in Section 2. Section 3 is dedicated to an overview of bootstrap techniques for Bayesian computation. Section 4 provides some theoretical insights and Section 5 describes the performance of the generative bootstrap sampler for Bayesian inference in some classical examples.



\subsection{Notation}
We denote $[n]$ as the set $\{1,2,\cdots,n\}$, $\bm 1_p\in\R^p$ as a vector of all 1's. We denote with $\|\cdot\|$ the Euclidean norm. For $X$, a random variable on a probability space $(\Omega,\Sigma,P)$, and $f$, a function on $\Omega$, we denote $\mathbb{P}(A)=\int_A dP$ for any $A\in\Sigma$, $\E[f(X)]=\int_\Omega fdP$, and also $Pf=\int_\Omega fdP$ to emphasize the underlying probability measure $P$.

\section{Setting the Stage}

Assume that we have observed a sequence of iid observations $\Xn=(x_1,\dots,x_n)'$ from an unknown sampling distribution $F_0$, i.e. $x_i \sim F_0$. 
 Bayesian inference traditionally requires the knowledge of the true underlying model for $F_0$.
This essentially boils down to specifying a family of likelihood functions $\mathcal F_\Theta=\{p_\theta^{(n)}(\Xn):\theta\in\Theta\}$ indexed by  an inferential  parameter $\theta\in\Theta\subseteq\R^p$. 
When there is uncertainty about the parametric family $\mathcal F_\Theta$ and mis-specification occurs, likelihood-based inference  can be misleading (\cite{kleijn2006misspecification})).
Without the obligation to construct a probabilistic model, it may often be easier to  infer about a target parameter $\theta$  that is directly tied to $F_0$, such as the mean, median or other quantile. In other words, one can merely express an interest in some functional of $F_0$ as opposed to parameter attached to a particular model $p_\theta^{(n)}(\Xn)=\prod_{i=1}^np_\theta(x_i)$.  Allowing  $\mathcal F_\Theta$ to be unknown, \cite{bissiri2016general} formalized a Bayesian framework for rational updating of beliefs by connecting data to parameters of interest via loss functions. 
We revisit their development in the context of optimization-centric Bayesian inference.

\subsection{Optimization-Centric Bayes}\label{sec:ocb}

Bayesian learning from experience and evidence typically processes information from two sources: (1) a prior density $\pi(\theta)$ which extracts  domain expertise, and (2) a likelihood function $p_\theta^{(n)}(\Xn)$ which distills the data.
Famously, the Bayes' rule   produces a post-data density summary for the parameters in a form of a posterior distribution $\pi(\theta\C\Xn)$.
There is, however,  a conceptually different path for arriving at  the posterior. Dating back to at least \cite{csiszar1975divergence} and \cite{donsker1975asymptotic}, Bayes' theorem can be interpreted as the 
optimal information processing rule solving  an infinite-dimensional optimization problem (\cite{zellner1988optimal}). Through the optimization-centric lens (\cite{knoblauch2019generalized}),   Bayesian  inference is viewed as regularized optimization
\begin{equation}\label{eq:opcentric}
\mathcal L(\ell; D;\Pi(\Theta))\equiv \arg\min_{q\in \Pi(\Theta)}\left\{\E_q\left[\sum_{i=1}^n \ell(\theta;x_i) \right] + D(q \| \pi)\right\}
\end{equation}
indexed by a triplet of parameters: (1) a loss function $\ell$,  (2) a discrepancy function $D(\cdot \| \cdot)$ gauging the departure from the prior, and  (3) a class of probability distributions $\Pi(\Theta)$  to optimize over.  The classical likelihood-based Bayesian inference yields 
\begin{equation}\label{eq:posterior_motivate}
\pi(\theta\C\Xn)=\mathcal L\left(-\log \pi_\theta(x); KL;\Pi(\Theta)\right)
\end{equation}
where $\Pi(\Theta)$ is unconstrained and where $KL$ stands for the Kullback-Leibler divergence.
The solution of the optimization problem in  \eqref{eq:posterior_motivate} can be rewritten more transparently as (see proof of Theorem 1 in \cite{knoblauch2019generalized} or \cite{csiszar1975divergence}) 
\begin{equation}\label{eq:insight}
\pi(\theta\C\Xn)=\arg\min_{q\in \Pi(\Theta)} KL\left[q\,\,\big\|\,\, \pi(\theta)\exp\left\{-\sum_{i=1}^n\ell(\theta;x_i)\right\}Z^{-1} \right],
\end{equation}
where $Z=\int_\theta\exp\{-\sum_{i=1}^n\ell(\theta;x_i)\}\pi(\theta)d\theta$ is the norming constant (assumed to be finite).
From \eqref{eq:insight} it can be seen that the optimization problem \eqref{eq:opcentric} forces priors and losses (log-likelihoods) into an exponentially additive relationship.
If the KL term was absent from \eqref{eq:posterior_motivate}, the solution would be a Dirac measure concentrated at the MLE estimator.
The incorporation of the prior $\pi(\theta)$ through the KL term in \eqref{eq:opcentric} allows one to obtain post-data densities for parameters in order to quantify uncertainty  and to perform belief updating. 
The formula \eqref{eq:insight} is a template for generating  belief  distributions  in more general situations, as will be seen below, through information theory.



The information-theoretic representation \eqref{eq:opcentric} reveals that committing to any particular Bayesian 
posterior is equivalent to committing to a particular optimization problem determined by (a) the loss function $\ell$, (b) the discrepancy metric $D$, and (c) the space of probability measures $\Pi(\Theta)$.
\cite{knoblauch2019generalized} presents various generalizations of Bayesian inference  by altering the parameters $(\ell; D; \Pi(\Theta))$ of the optimization objective in \eqref{eq:opcentric}.
For example, constraining the class of distributions $\Pi(\Theta)$ to some parametric form  is equivalent to the variational Bayes approach (\cite{wainwright2008graphical}). 
Alternatively, replacing the self-information loss $\ell(\theta;x)=-\log\pi_\theta(x)$ with any other loss function, one obtains the so called Gibbs posteriors (more below).
Following \cite{knoblauch2019generalized}, we regard \eqref{eq:opcentric} as the unifying hat behind various generalized Bayesian inference methods.

\subsection{Bayesian Inference with Gibbs Posteriors}\label{sec:loss}
Sometimes, the parameter interest $\theta$ is defined indirectly through a general loss function $\ell(\theta;x)$ rather than the likelihood function.
For an unknown distribution function $F_0$,   from which we observe an iid vector $\Xn$, one can define such inferential target $\theta_0$
as the minimizer of the average loss \citep{bissiri2016general}
\begin{equation}\label{eq:theta0}
\theta_0= \theta(F_0)\equiv\arg\min\limits_{\theta\in\Theta}\int\ell(\theta;x)dF_0(x).
\end{equation}
Replacing $F_0$ with the empirical distribution function  $P_n$ of $\Xn$, one obtains an empirical risk minimizer, for example an M-estimator \citep{huber2004robust,{maronna2006robust},{huber2009robust}}. In econometrics, extremum estimators   (e.g. censored quantile regression of \cite{powell1986censored} or instrumental and robust median regression \cite{mood1950introduction})  are also defined as maximizers of a random finite-sample criterion function whose population counterpart is maximized uniquely at some point $\theta_0\in\Theta$. 
In these examples,  $\theta_0$ generally cannot be understood as a model parameter but rather  as a solution to   an optimization (loss minimization) problem.
\cite{chernozhukov2003mcmc} note that implementing such estimators can be challenging and introduce quasi-Bayesian estimators for estimation and inference. 
\cite{bissiri2016general} propose a related general framework for updating belief distributions as a Bayesian extension of M-estimation.
Indeed, even when the parameter cannot be directly assigned any particular model interpretation, it is still possible   to perform Bayesian belief updating and uncertainty quantification. 

In particular, \cite{bissiri2016general} suggest a decision-theoretic representation of beliefs about $\theta$ via a composite loss function over probability measures which gauges fidelity to data and departure from the prior (\cite{berger1993statistical}).
Curiously, their loss function corresponds to the optimization objective in \eqref{eq:opcentric} where the log-likelihood has been replaced by a general loss function $\ell(\theta;x)$ and where $D(\cdot\|\cdot)$ is the KL divergence.
Similarly as   in Section \ref{sec:ocb}, it can be shown that the optimal distribution which minimizes this cumulative loss function (without constraining $\Pi(\Theta)$) has an exponentially additive form
\begin{equation}\label{eq:gibbs}
\tilde \pi(\theta\C\Xn)= \frac{\pi(\theta)\exp\left\{-\alpha \sum_{i=1}^n\ell(\theta;x_i)\right\}}{\int_{\Theta}\pi(\theta)\exp\left\{-\alpha\sum_{i=1}^n\ell(\theta;x_i)\right\}d\theta},
\end{equation}
where $\alpha=1$. Other values of $\alpha>0$ have been considered  to regulate the speed of  the learning rate (see \cite{holmes2017assigning,grunwald2012safe}).
The distribution \eqref{eq:gibbs}  is the ``quasi-Bayesian'' posterior introduced in \cite{chernozhukov2003mcmc} and it became known as the Gibbs posterior,  a probability distribution for random estimators defined by an empirical measure of risk (\cite{catoni2004statistical,zhang2006information}). 
The inferential object \eqref{eq:gibbs}  now does not have the interpretation of the usual posterior but rather an optimal prior-to-posterior updating distribution that satisfies coherence and information preservation requirements.


The motivation for Gibbs posteriors can be traced back to the early work of  Laplace (\cite{laplace1774memoire}) who regarded a transformation of a least square criterion function as a statistical belief and obtained point estimates of that distribution ``without any assumption about the error distribution'' (\cite{stigler1975studies}).
 In thermodynamics, the risk is interpreted as an energy function. 
In the PAC-Bayesian approach (\cite{shawe1997pac,McAllester1998somepac-bayesian}), the Gibbs distribution appears as the probability distribution that minimizes the upper bound of an oracle inequality on the risk of   estimators.
Estimators derived from Gibbs posteriors, such as quasi-Bayesian mean or median \citep{chernozhukov2003mcmc}, usually show excellent performance and yet their actual implementation can be challenging. The usual recommendation (\cite{dalalyan2012mirror,alquier2013sparse}) is to sample from a Gibbs posterior using MCMC (see e.g. \cite{green2015bayesian} or \cite{ridgway2014pac} who propose tempering sequential Monte Carlo which may be too slow for practical use). \cite{alquier2016properties} propose a variational Bayes approximation which can achieve the same rate of convergence as the actual Gibbs posterior 
and  which has  a polynomial time complexity in convex problems. Our work explores Bootstrap techniques and generative bootstrap samplers for approximating Gibbs posteriors, going beyond the development in \cite{lyddon2019general}.
Before delving into the implementation aspects,  we distinguish the parametric inference approach with Gibbs posteriors (\cite{chernozhukov2003mcmc,bissiri2016general})  from the non-parametric Bayesian learning approach (\cite{chamberlain2003nonparametric,lyddon2019general}). 

\subsection{Bayesian Non-parametric Learning}\label{sec:nbl}
Gibbs posteriors lock priors and losses in an exponentially additive relationship in order to achieve coherent Bayesian updating of beliefs. 
The uncertainty about the inferential target is a-priori represented in the prior distribution $\pi(\theta)$. The Bayesian non-parametric learning (NPL)  approach (\cite{lyddon2019general,chamberlain1996nonparametric}), on the other hand,
 expresses uncertainty about $\theta$ through a prior on the unknown distribution function $F_0$.

Defining the parameter of interest as a functional of $F_0$ (as in \eqref{eq:theta0}), the focus is shifted from $\theta_0$ to $F_0$. 
\cite{lyddon2019general} propose a two-step Bayesian learning process by assigning a Dirichlet process (DP)  prior  $F\sim DP(\alpha, F_\pi)$ on the unknown distribution function $F$. 
The base measure $F_\pi$ conveys prior knowledge about the sampling distribution $F_0$ and, indirectly, also the parameter $\theta_0$.
In the first step, Bayesian non-parametric learning is used to form beliefs about the joint nonparametric density of data and then draws of the non-parametric density are made to repeatedly compute the extremum parameter of interest.  \cite{lyddon2019general} call this sampling procedure the loss-likelihood bootstrap. \cite{chamberlain1996nonparametric} first introduced this strategy using a particular DP posterior (supported only on observations $\Xn$ with Dirichlet-distributed probabilities for each state) that corresponds to the Bayesian bootstrap (\cite{efron1979bootstrap}).

 A suitable choice for the base measure $F_\pi$ is the empirical distribution of the historical data. The second component of the DP prior the concentration parameter $\alpha>0$ which can be re-interpreted as the effective sample size from the prior $F_\pi$. 
Assigning a DP prior on $F_0$, the posterior $F\C\Xn\sim DP(\alpha+n, G_n)$ is also DP with the base measure updated as $G_n=\frac{\alpha}{\alpha+n}F_\pi+\frac{1}{\alpha+n}\sum_{j=1}^n\delta_{x_i}$. The posterior for the inferential target $\theta$ is then determined from this posterior through the mapping \eqref{eq:theta0}. In particular, under the stick-breaking representation  (\cite{sethuraman1994constructive}) of the DP posterior, draws from $F\C \Xn$ are almost surely discrete and the parameter of interest, for each $F$, can be computed as
$$
\theta(F)=\arg\min_\theta\sum_{k=1}^\infty w_k \times \ell(\theta;y_k)
$$ 
where $w_k$'s are the stick-breaking beta-products and where $y_k$ are iid samples from $G_n$. Drawing $F$ from the DP posterior  requires infinite time when $F_\pi$ is continuous. \cite{fong2019scalable} suggest  an approximate sampling scheme based on a truncation approximation of $F_\pi$ and bootstrap-style sampling. The idea is to generate  $T$ fake data points $\tilde x_j$ from $F_\pi$ and assign each one a random weight $\tilde w_j$. The weights $\bm{w}=(w_1,\dots, w_n)'$ for the observed data and fake data $\tilde{\bm{w}}=(\tilde w_1,\dots,\tilde w_T)'$ are jointly sampled from  a Dirichlet distribution $Dir(1,\dots, 1,\alpha/T,\dots,\alpha/T)'$. The posterior sample of the parameter $\theta(F)$ is then obtained by minimizing an augmented objective $\sum_{i=1}^n w_i \ell(x_i;\theta)+\sum_{j=1}^T\tilde w_j\ell(\tilde x_j;\theta)$. Bayesian bootstrap is obtained as a special case when $\alpha=0$.

The non-parametric Bayesian learning approach (with Bayesian bootstrap (\cite{efron1979bootstrap}) or the loss-likelihood bootstrap (\cite{lyddon2019general})) requires numerous re-computations of the extremum estimates in order to construct the posterior distribution over the parameter of interest. This can be prohibitively slow for optimization problems that are costly to solve.  In this work, we investigate the possibility of  deploying  the generative Bootstrap sampler of \cite{shin2020scalable} that learns a deterministic mapping of weights $(\bm w,\tilde{\bm w})$ to obtain samples of $\theta(F)$.

\section{Bootstrap for Bayesian Computation}
Sampling methods are innate to Bayesian computation. Parallelizable iid sampling (with Approximate Bayesian Computation (\cite{pritchard1999population,beaumont2002approximate}) or bootstrap (\cite{newton1994approximate,newton2021weighted}) has certain privileges over sequential sampling with MCMC.
ABC techniques are useful when the likelihood is easier to sample from than to evaluate. Bootstrap-style samplers (\cite{newton1994approximate,rubin1981bayesian,efron2012bayesian,newton2021weighted,nie2020bayesian,shin2020scalable}), on the other hand,  are beneficial when optimization is easier than, for example, sampling from conditionals. 
Below, we review recent developments in  bootstrap-style posterior computation  for Bayesian inference with loss functions.

Weighted likelihood bootstrap (WLB) of \cite{newton1994approximate} is a method for approximately sampling from a posterior distribution of a well-specified parametric statistical model. Samples are generated by computing randomly-weighted maximum likelihood estimates with the weights drawn from a suitable Dirichlet distribution.  
Such bootstrap samples are obtained as minimizers of randomly re-weighted objective functions, e.g. 
\begin{equation}\label{eq:theta_hat}
\wh \theta_{\bm w}=\arg\min_\theta \left\{\sum_{i=1}^n w_i \times \ell(\theta;x_i)-\lambda\log\pi(\theta)\right\}
\end{equation}
for a given set of weights $\bm w\in\R^n$ drawn from some distribution $H$ and  for  the log-likelihood loss function $\ell(\cdot;x_i)$. 
The WLB method  is not an exact method and does not accommodate a prior (i.e. \cite{newton1994approximate} assume $\lambda=0$). \cite{newton2021weighted} added a log-prior penalty  to incorporate the prior with $\lambda>0$.
Conceivably, one can consider any general loss function $l(\cdot;y_i)$ and use alike strategy to sample from approximate Gibbs posteriors. We illustrate this in a Bayesian least absolute deviation regression example in Section \ref{sec:LAD}. 

The WLB sampler can be in principle used to  construct approximate quasi-posteriors (\cite{chernozhukov2003mcmc}), where the prior may need to be incorporated in some form either through distribution $H$ or through a penalty term in \eqref{eq:theta_hat}). 
This aims at parametric-type inference with Gibbs posteriors (as described in Section \ref{sec:loss}).
Loss-likelihood bootstrap (\cite{lyddon2019general}) reinterprets WLB as a sampler from an exact posterior over a parameter of interest defined through a loss function under an unknown sampling distribution. 
This aims at non-parametric Bayesian learning (as described in Section \ref{sec:nbl}).
This is also the strategy pursued in \cite{chamberlain2003nonparametric} based on the Bayesian bootstrap (\cite{rubin1981bayesian}).

\subsection{Deep Bootstrap Sampling}\label{sec:deep_bootstrap_sampling}
Recall that the loss-likelihood bootstrap of \cite{lyddon2019general} generates samples 
\begin{equation}
\wh\theta_{\bm w}=\theta(F_j)\quad \text{where}\quad F_j=\sum_{i=1}^n\delta_iw_i\label{eq:thetaw}
\end{equation}
with $w_i$'s arriving from, e.g.,   a Dirichlet distribution and where $\wh\theta_{\bm w}$ can be viewed as a minimizer of an expected loss under $F_j$ as defined in \eqref{eq:theta0}. 
At a more intuitive level, each $\wh \theta_{\bm w}$  in \eqref{eq:theta_hat} can   be seen as a flexible functional of $\bm w$. 
This suggests a compelling possibility of  treating the distribution of  $\wh\theta_{\bm w}$ (be it an approximation to the Gibbs posterior or the non-parametric Bayesian posterior)
as an implicit distribution. A distribution is implicit when it is not possible to evaluate its density but it is possible to draw samples from it. One typical way to draw from an implicit distribution is to first sample a noise vector and then push it through a deep neural network (\cite{mohamed2016learning}). Implicit distributions have been deployed within variational Bayes (\cite{ruiz2019contrastive,pequignot2020implicit}) to obtain flexible distributional approximations to the posterior. Treating bootstrap distributions implicitly,
the generative bootstrap sampler (\cite{shin2020scalable})  draws  samples by learning a flexible mapping which transports weights $\bm w$ onto parameters $\theta(F)$.  A similar strategy can be used for the WLB approach of \cite{newton1994approximate} where $\wh\theta_{\bm w}$ is linked to $\bm w$ through \eqref{eq:theta_hat}.

Instead of re-computing the optimization problem \eqref{eq:theta_hat} for a freshly drawn set of weights $\bm w$ at each step, \cite{shin2020scalable} suggest training a generator mapping, say $\wh \theta(\bm w)$, 
which has to be learned only once. This mapping is designed to pass random weights $w_j$'s to yield samples from the bootstrap distribution, a sampling process which has negligible cost once the mapping has been learned.
 The following Lemma (Theorem 2.1 in  \cite{shin2020scalable}) justifies this line of reasoning.


\begin{lemma}(\cite{shin2020scalable})\label{lemma1}
Assume that a function $G:\R^n\rightarrow\R^p$ is defined as
\begin{equation}\label{eq:Ghat}
 \hat G(\cdot)=\arg\min_{G\in\mathcal G} \E_{\bm w}\sum_{i=1}^n w_i \times \ell(G(\bm w); y_i)
\end{equation}
where $\E_{\bm w}$ is the expectation with respect to $\bm w\sim H$.  Moreover, assume that the solution $\wh\theta(\bw)$ defined in \eqref{eq:theta_hat} is unique for each given $\bm w\in\R^n$. Then
if $\mathcal G$ is rich enough to express any function $G$ we have
\begin{equation}\label{eq:ghat}
\wh G(\bm w)=\wh \theta_{\bm w}.
\end{equation}
\end{lemma}
\proof 

The proof is given in Section 2.2 of \cite{shin2020scalable} and rests on the simple observation that, since
$\sum w_il(\theta;y_i)\geq \sum {w_i}l(\wh\theta_{\bw};y_i),$
one has
$$
 \E_{\bm w}\sum_{i=1}^n w_i l(\wh G(\bm w); y_i)\geq  \E_{\bm w}\sum_{i=1}^n w_i l(\wh\theta_{\bw}; y_i)
$$
which implies that  $\wh G(\bw)=\wh\theta_{\bw}.$

The result in Lemma \ref{lemma1} implies an important ``isomorphism'' between bootstrap weights $\bm w$ and parameters $\theta$ which can be exploited for faster computation of belief distributions (Gibbs posteriors in Section \ref{sec:loss}) or non-parametric Bayesian learning posteriors (Section \ref{sec:nbl}). 
For training $G(\cdot)$, one may want to search within mappings $\mathcal G$ that are compositions of  non-linear transformations, i.e. a deep learning mappings. 
Due to the  expressibility of neural networks (see e.g. \cite{barron1993universal})), the neural network estimator $\hat G(\cdot)$ can be made arbitrarily close to the optimal mapping that satisfies \eqref{eq:ghat}. 
This work uses forward deep learning mappings $\mathcal G$, thereby the name Deep Bootstrap Sampler.

\begin{figure}[t]\centering
\includegraphics[width=.75\linewidth, height=.23\textheight]{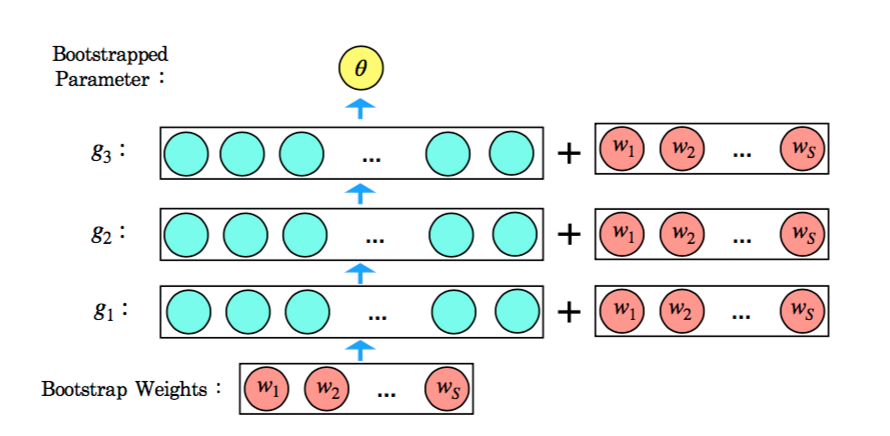}
\caption{Prototype deep learning architecture from \cite{shin2020scalable}.}
\label{fig:architecture}
\end{figure}

The nested structure of neural networks allows for gradients to be efficiently evaluated using back-propagation (\cite{hecht1992theory,rumelhart1986learning}). Once the gradient is computed, stochastic gradient descent algorithm can be used to update the parameters iteratively. \cite{shin2020scalable} also suggest a specific neural network architecture which re-introduces the weights at each layer.
Such a deep approximating class with $L$ network layers, each with $n_l$ neurons,  can be written as 
\begin{align*}
g(\bm w)&=g_L(\bm Z_L)\\
\bm Z_{l+1}&=\{g_l(\bm Z_l),\bm w\}\quad \text{for}\quad 1\leq l\leq L-1.
\end{align*}
where $\bm Z_1=\bm w$ and where each function $g_l:\R^{n_l}\rightarrow \R^{n_{l+1}}$ squashes a multivariate linear combination of the input variables $Z_l$ by a squashing function $T$, e.g. the Rectified Linear Unit (ReLU) or a sigmoid function. The parameters of the linear combinations for each layer (i.e. intercept terms and slopes) are encapsulated in a vector parameter $\bm\phi$.

{In the neural network architecture proposed by \cite{shin2020scalable}, the number of trainable parameters in the $l$-th layer is $O((n_l+n)n_l)$. When $n\gg n_l$, this number grows linearly in $n$ and thus, reducing the dimension $n$ of input will significantly improve the computational efficiency.  \cite{shin2020scalable} proposed a subgroup bootstrap strategy, which groups the $n$ weights into $S$ equi-sized blocks and $S\ll n$. The weights in the same group are assigned the same value. The $S$ weights are drawn from some distribution $H_\alpha$. Then the number of trainable parameters in the $l$-th layer will be reduced to  $O((n_l+S)n_l)$. The subgroup strategy significantly boosts the computational advantage of Deep Bootstrap samplers. The architecture incorporating this strategy is shown in Figure \ref{fig:architecture}. }
We summarize the Deep Bootstrap sampler for Bayesian parametric learning with Gibbs posteriors in Algorithm \ref{alg:gibbs} and  for Bayesian non-parametric learning in Algorithm \ref{alg:nonparam}. 

	{\begin{algorithm}[t]\scriptsize
		\spacingset{0.7}
		\KwData{Data $\{x_i,1\leq i\leq n\}$, number of training epochs $T$, number of points to sample $N$, number of Monte Carlo samples $K$, prior $\pi$, learning rate $\eta$, number of subgroups $S$ for observed data, $n_S=n/S$ the size of each subgroup.}
		\KwResult{$\theta^i, i=1,2,\cdots,N$.}
		\hspace{-0.005\linewidth}\colorbox{gray!30}{\makebox[0.99\linewidth][l]{Training stage:}}
		
		\textbf{Initialize weights $\phi$ of the fitted function $\hat G$.}
		
		\For{$t=1,2,\cdots, T$}{
			\For{$k=1,2,\cdots,K$}{
				{\vspace{-0.3cm} \hspace{5cm}\color{white} \tiny ahoj}\\
				\textbf{Draw $\tilde w^k_{1:S}\sim S\times Dir(1,\cdots,1)$}.
				
				\textbf{Set} $w_{(j-1)n_S+m}^k=\tilde w_j^k$ for all $m=1,2,\cdots,n_S$, $j=1,2,\cdots,S$.
			}
			
			\textbf{Update} $\phi\leftarrow\phi-\eta\partial_{\phi}\left[\sum_{k=1}^K\left[\sum_{j=1}^{n} w_j^kl\left(G(\tilde w^k_{1:S}); x_j\right)+\log\pi\left(G(\tilde w^k_{1:S})\right)\right]\right]$.
			
		}
	
		\hspace{-0.005\linewidth}\colorbox{gray!30}{\makebox[0.99\linewidth][l]{Sampling stage:}}
		
		\For{$i=1,2,\cdots,N$}{
				{\vspace{-0.3cm} \hspace{5cm}\color{white} \tiny ahoj}\\
			\textbf{(2.1)} Draw $\tilde w^k_{1:S}\sim S\times Dir(1,\cdots,1)$.\\
			\textbf{(2.2)} Evaluate $\theta^i=\hat G_\phi(\tilde w^k_{1:S})$.
		}
		\caption{\bf : Deep Bootstrap Sampler for Gibbs Posterior Learning}\label{alg:gibbs}
\end{algorithm}
}

{
	\begin{algorithm}[t]\scriptsize
		\spacingset{0.7}
		\KwData{Data $\{x_i,1\leq i\leq n\}$, number of prior pseudo samples $n'$, number of training epochs $T$, number of points to sample $N$, number of Monte Carlo samples $K$, concentration parameter $\alpha$, learning rate $\eta$, number of subgroups $S$ for observed data, number of subgroups $S'$ for prior pseudo data, $n_S=n/S, n_{S'}=n/S'$.}
		\KwResult{$\theta^i, i=1,2,\cdots,N$.}
		\hspace{-0.005\linewidth}\colorbox{gray!30}{\makebox[0.99\linewidth][l]{Training stage:}}
		
		\textbf{Initialize weights $\phi$ of the fitted function $\hat G$.}
		
		\textbf{Approximate the prior} by drawing ${x}'_{1:n'}\stackrel{\iid}{\sim}F_\pi$. 

		\textbf{Create} the enlarged (observed + prior pseudo) sample $\{x_1,\cdots,x_n,x_1',\cdots,x_{n'}'\}$.
		
		\For{$t=1,2,\cdots, T$}{
			\For{$k=1,2,\cdots,K$}{
				{\vspace{-0.3cm} \hspace{5cm}\color{white} \tiny ahoj}\\
				\textbf{Draw $\tilde w^k_{1:(S+S')}\sim (S+S')\times Dir(1,\cdots,1,\alpha/n',\cdots,\alpha/n')$.}
				
				\textbf{Set} $w_{(j-1)n_S+m}^k=\tilde w_j^k$ for any $m\in[n_S],\,j\in[S]$ and $w_{n+(j'-1)n_{S'}+m'}^k=\tilde w_{j'+S}^k,$ for any $m'\in[n_{S'}], j'\in[S']$.
			}

			\textbf{Update} $\phi\leftarrow\phi-\eta\partial_{\phi}\left[\sum_{k=1}^K\left[\sum_{j=1}^{n} w_j^kl\left(G(\tilde w^k_{1:(S+S')}); x_j\right)+\sum_{j=1}^{n'} w_{j+n}^kl\left(G(\tilde w^k_{1:(S+S')}k); x'_j\right)\right]\right]$.
			
		}
		
		\hspace{-0.005\linewidth}\colorbox{gray!30}{\makebox[0.99\linewidth][l]{Sampling stage:}}
		
		\For{$i=1,2,\cdots,N$}{
			{\vspace{-0.3cm} \hspace{5cm}\color{white} \tiny ahoj}\\
			\textbf{(2.1)} Draw $\tilde w^k_{1:(S+S')}\sim (S+S')\times Dir(1,\cdots,1,\alpha/n',\cdots,\alpha/n')$.\\
			\textbf{(2.2)} Evaluate $\theta^i=\hat G_\phi(\tilde w^k_{1:(S+S')})$.
		}
		\caption{\bf : Deep Bootstrap Sampler for Bayesian NPL}\label{alg:nonparam}
\end{algorithm}
}

\section{Theory}
The quantification of the speed of concentration around the truth (as the sample size goes to infinity) is now a standard way of assessing the quality of posteriors (\cite{ghosal2000convergence}).
As \cite{shalizi2009dynamics} states, such Bayesian asymptotic results are ``frequentist license for Bayesian practice''. Below, we review recent literature related to our development and provide theoretical support for certain aspects of the weighted likelihood bootstrap  using connections to model misspecification.

\cite{bhattacharya2019bayesian} studied concentration of the so-called fractional $\alpha$-posteriors obtained by raising the likelihood to some fixed value $\alpha\in(0,1)$. We study bootstrap-style posteriors where each observation is raised 
to a {\em different random} weight $w_i$ where $\sum_{i=1}^nw_i=n$.  {\cite{bhattacharya2019bayesian} proved that the fractional $\alpha$-posteriors concentrate on the so-called $\alpha$-divergence neighborhoods around the truth. The $\alpha$-divergence is shown to be a valid divergence measure when $\alpha\in(0,1)$, and the rate of contraction is inflated by a multiplicative factor $\frac{1}{1-\alpha}$. This line of proof, unfortunately, does not extend easily to our bootstrap-style posteriors.\footnote{Unless $w_i=1$ for all $i$'s, there must be some weight $w_i> 1$. The existence of such weights invalidates the upper bound in \cite{bhattacharya2019bayesian}, and we find it difficult to define a similar (and valid) divergence measure that could be properly upper bounded.}} We study the distribution of extremal (modal)  estimators  
$\wh\theta_{\bm w}$ defined in \eqref{eq:theta_hat} when $\lambda=1$.
While the $\alpha$-posteriors keep the weight fixed and the randomness stems from treating $\theta$ as a random variable with a prior, we treat $\hat\theta_{\bm w}$ as an estimator where the randomness comes from $\bm w$. In a related paper, \cite{han2019statistical} study  contraction of weighted posterior distributions incorporating both the randomness of $\theta$ (through the weighted posterior distribution under the prior $\pi(\theta)$) and the randomness of $\bm w$ (from the distribution of weights $\pi(\bm w)$). 
Our theory has two parts. First, in our Theorem \ref{thm:concentration},  we obtain a similar conclusion to \cite{han2019statistical} but using a different proving technique. There we focus on the entire weighted posteriors assuming that $\ell(x;\theta)$ in \eqref{eq:theta_hat} is the unit information loss. We refer to \cite{syring2020gibbs} for concentration-rate results for the actual Gibbs posteriors with sub-exponential loss functions. 
Second, we are  interested in the contraction  of a distribution of weighted posterior modes $\hat\theta_{\bm w}$ around $\theta_0$. This  property is summarized in Theorem  \ref{thm:mode}, where we consider  losses other than just the unit information loss.

First, we want to understand the behavior of the weighted posteriors 
\[
\pi_{\bm w}(\theta\mid\Xn)=\frac{\pi(\theta)p_{\theta}^{(n),\bm w}(\Xn)}{\int_{\Theta}\pi(\theta)p_{\theta}^{(n),\bm w}(\Xn)d\theta}\propto e^{u_{\theta,\bm w}(\Xn)}p_{\theta}^{(n)}(\Xn)\pi(\theta),
\]
where the exponential tilting factor
$$
u_{\theta,\bm w}(\Xn)=\log\frac{p_{\theta}^{(n),\bm w}(\Xn)p_{\theta_{0}}^{(n),1-\bm w}(\Xn)}{p_{\theta}^{(n)}(\Xn)}
$$
combines individually weighted likelihood terms (with weights $w_i$ and $1-w_i$)
$$
p_{\theta}^{(n),\bm w}(\Xn)=\prod_{i=1}^np_{\theta}^{w_i}(X_i)\quad\text{and}\quad
p_{\theta_0}^{(n),1-\bm w}(\Xn)=\prod_{i=1}^np_{\theta_0}^{1-w_i}(X_i).
$$
Following \cite{kaji2021metropolis}, for any fixed $\bm w$, $\pi_{\bm w}(\theta\mid\Xn)$ can be viewed as the posterior density under a mis-specified likelihood
\[
\wt{p}_{\theta}^{(n),\w}(\Xn)=\frac{e^{u_{\theta,\bm w}(\Xn)}p_{\theta}^{(n)}(\Xn)}{C_{\theta,\bm w}}\quad\text{where}\quad C_{\theta,\bm w}=\int_{\mathcal{X}}e^{u_{\theta,\bm w}(\Xn)}p_{\theta}^{(n)}(\Xn)d\Xn,
\]
and a modified prior 
\begin{equation}\label{eq:modified_prior}
\wt{\pi}_{\bm w}(\theta)=\frac{C_{\theta,\bm w}\times \pi(\theta)}{\int_{\Theta}C_{\theta,\bm w}\times \pi(\theta)}.
\end{equation}
Note that $P_{\theta_0}^{(n)}$ is \emph{not} of the same form as $\mathcal{\wt{P}}^{(n),\bm w}=\{\tilde P_\theta^{(n),\bm w}:\theta\in\Theta\}$. This new mis-specification perspective allows us to use a different proving technique from \cite{bhattacharya2019bayesian} and \cite{han2019statistical}.
It is known (\cite{kleijn2006misspecification}) that under the mis-specified model $\wt{p}_{\theta}^{(n),\w}$, the posterior will concentrate around the KL-projection point $\theta_{\bm w}^*$ defined as
\begin{equation}\label{eq:dfn_theta_W^*}
\theta_{\bm w}^*=\arg\min_{\theta\in\Theta}-P_{\theta_0}^{(n)}\log\frac{\wt{p}_{\theta}^{(n),\w}(\Xn)}{p_{\theta_0}^{(n)}(\Xn)}.
\end{equation}
At first sight, since the model $\wt{p}_{\theta}^{(n),\w}$ is mis-specified and depending on the value of $\bm w$,  $\theta_{\bm w}^*$ will not necessarily be the true target $\theta_0$. Surprisingly, under some mild conditions on $\bm w$, $\theta_0$ will be the unique minimizer of Equation \eqref{eq:dfn_theta_W^*}, i.e., $\theta_0=\theta_{\bm w}^*$. 
This is stated in the next Lemma.

\begin{lem}\label{lemma:unique_minimizer}
	Assume that: (1) 
	there exists $i\in[n]$ such that $w_i>0$,  and (2) the random variable $Z(X_i)=\frac{p_{\theta}\left(X_{i}\right)}{p_{\theta_0}\left(X_{i}\right)}$ is degenerate only if $\theta=\theta_0$. 
	Then 
	\begin{equation*}
	\theta_0=\arg\min_{\theta\in\Theta}-P_{\theta_0}^{(n)}\log\frac{\wt{p}_{\theta}^{(n),\w}(\Xn)}{p_{\theta_0}^{(n)}(\Xn)},\quad\text{i.e.},\quad \theta_{\bm w}^*=\theta_0.
	\end{equation*}
\end{lem}
\proof
See Appendix Section \ref{sec_appendix:unique_minimizer_proof}.

The lemma says that even though we perturb the likelihood, the truth $\theta_{0}$ still
optimizes the KL divergence between the mis-specified likelihood $\wt{p}_{\theta}^{(n),\w}$
and the true likelihood $p_{\theta_{0}}^{(n)}$. 
{The intuitive explanation
is that 
\[
\wt{p}_{\theta}^{(n),\w}=C^{-1}_{\theta,\w}p_{\theta_{0}}^{(n),1-\w}p_{\theta}^{(n),\w}.
\]
The mis-specified likelihood $\wt{p}_{\theta}^{(n),\w}$
can be thereby viewed as the original likelihood $p_{\theta_{0}}^{(n)}$ adjusted
towards $p_{\theta}^{(n)}$, where the strength of adjustment depends
on  $\bm w$. When $\theta=\theta_{0}$, we always have $\wt{p}_{\theta}^{(n),\w}=p_{\theta_{0}}^{(n)}$,
regardless of the weights $\bm w$. }
To start, using  Lemma \ref{lemma:unique_minimizer}, we can conclude concentration-rate result for the whole posterior $\pi_{\bm w}(\theta\mid X^{(n)})$ around the truth $\theta_0$. 
Define  $R_n(c_0)=\{(a_1,a_2,\cdots,a_n)'\in\mathbb{R}^n: a_i\ge 0, \sum_{i=1}^n a_i=n, \max_{i\in[n]} a_i\le c_0\log n\}$.

\begin{thm}\label{thm:concentration}
	Assume that there exists a constant $c_0>0$ and
	a sequence of real numbers $\epsilon_{n}>0$ with
	$\epsilon_{n}\rightarrow0,n\epsilon_{n}^{2}\rightarrow\infty$, such that the  sequence of sets $R_n(c_0)$ 
	satisfies
	\begin{equation}\label{eq:weight_regularization}
	\mathbb{P}_{\bm w}[\w_n\in R_n(c_0)]\rightarrow 1.
	\end{equation}
	Under technical Assumptions \ref{assump_appendix:metric}, \ref{assump_appendix:test},  \ref{assump_appendix:prior} in the Appendix (Section \ref{sec_appendix:concentration_proof}) and the assumptions in Lemma \ref{lemma:unique_minimizer} we have 
	\[
	P_{\theta_{0}}^{(n)}P_{\bm w}\left[\Pi_{\boldsymbol{w}}\left(\left\Vert \theta-\theta_{0}\right\Vert >M_{n}\epsilon_{n}\mid X^{(n)}\right)\right]\rightarrow0
	\]
	for any sequence $M_{n}\rightarrow\infty$ as $n\rightarrow\infty$, where $\Pi_{\bm w}(\cdot\mid X^{(n)})$ is the probability measure associated with the posterior density $\pi_{\bm w}(\cdot\mid X^{(n)})$.
\end{thm}
\proof See Appendix Section \ref{sec_appendix:concentration_proof}.

\cite{han2019statistical} obtained a similar result  under stronger differentiability assumptions on both the prior and the likelihood to achieve a parametric rate (up to a logarithmic factor).  Our Theorem \ref{thm:concentration}  assumes a prior mass and testing conditions, yielding a more general concentration rate (not necessarily $1/\sqrt{n}$). In addition, our proving technique is different and uses the model mis-specification perspective.

We are ultimately interested in the convergence of posterior modes $\hat\theta_{\bm w}$ around $\theta_0$ where the randomness is driven by $\bm w$. 
{Theorem \ref{thm:concentration} implies the existence of a point estimator based on $\Pi_{\bm w}(\cdot\mid X^{(n)})$ that converges to the truth at a rate $\epsilon_n$ for each $\bm w$ (e.g., using Theorem 8.7 in \cite{ghosal2017fundamentals}). 
In  Theorem 8.8 of \cite{ghosal2017fundamentals}, for example, it is shown that for convex distances, the posterior mean converges at a rate $\epsilon_n$.
The convergence rate might be adopted also by the posterior mode under suitable local asymptotic normality conditions. We have chosen a different, more direct, route to show the convergence of posterior modes  $\hat\theta_{\bm w}$.
}
 We utilize tools for establishing convergence rates for M-estimators (\cite{wellner2013weak}). {We denote with  $p_\theta(x)=\exp\{-l(\theta;x)\}$ an exponentiated loss, not necessarily a likelihood, and we  show that $\hat\theta_{\bm w}$ concentrates around  the inferential target $\theta_0$  defined in Equation \eqref{eq:theta0}.

}

We denote $\mathcal{M}_\epsilon(\theta_0)=\left\{\log[p_{\theta}/p_{\theta_0}]:d(\theta,\theta_0)<\epsilon\right\}$, {and $P_0$   the probability measure of $X_i$'s, i.e., $X_i\stackrel{\text{iid}}{\sim} P_0$.}
For any function class $\mathcal{F}$, we write its Rademacher complexity with respect to $P_{0}$ for the sample size $n$ as $\mathcal{R}_n(\F)$, i.e.,
\[
\mathcal{R}_n(\F)=P^{(n)}_{0}\left[\sup_{f\in\F}\left|\frac{1}{n}\sum_{i=1}^{n}\sigma_if(X_i)\right|\right],
\]
where $\P(\sigma_i=1)=\P(\sigma_i=-1)=1/2$ for any $i\in[n]$.
Denote $\phi_n(\epsilon)=\sqrt{n}\mathcal{R}_n(\mathcal{M}_\epsilon(\theta_0))$.
The following theorem establishes the convergence rate of the posterior modes $\hat\theta_{\bm w}$. We allow the prior $\pi$ to depend on the sample size $n$, and we thereby write $\pi_n$ for emphasis.

\begin{thm}\label{thm:mode}
	Assume that:
	\begin{itemize}
		
		\item[(1)] (Existence of a suitable semi-metric) There exists a semi-metric $d(\cdot,\cdot)$ on $\Theta$ such that \[P_0\log [p_{\theta_0}/p_{\theta}]\ge d^2(\theta,\theta_0)\quad\text{for all}\quad \theta\in\Theta,\] 
		
		\item[(2)] (Bounded loss)  $\sup_{\theta\in\Theta}\left|\log p_{\theta}(X)\right|<\infty$,
		
		\item[(3)] (Weight regularization) The weights satisfy
		\begin{equation}
		w_i\,\text{i.i.d. with}\,\,\E w_i=1, \|w_i\|_{2,1}=\int_0^\infty\sqrt{P\left(|w_i|>x\right)}dx<\infty,\quad\text{or}
		\end{equation}
		\begin{equation}
		\bm w_n=(w_1,\cdots,w_n)\sim n\times Dir(c,\cdots,c),\quad\text{for some fixed }c>0.
		\end{equation} 
		
		\item[(4)] (Proper growth of Rademacher complexity) There exist constants $C>0$, $\gamma\in(0,2)$ such that for all $\epsilon>0,c>1$, 
		\[
		\phi_n(\epsilon)\le C\sqrt{n}\epsilon^2,\quad
		\phi_n(c\epsilon)\le c^{\gamma}\phi_n(\epsilon).
		\]
	\end{itemize}
	Then, for any $M_n\rightarrow\infty$  as $n\rightarrow\infty$, we have
	\[
	P_{0}^{(n)}\P_{\bm w}\left(d(\hat\theta_{\bm w},\theta_0)>M_n\epsilon_{n}\right)\rightarrow 0,
	\]
	where $\epsilon_{n}\rightarrow 0$ satisfies 
	\begin{equation}\label{eq:mode_rate}
		\epsilon_{n}^{-2}\phi_n(\epsilon_{n})\le\sqrt{n}\,\,\text{for all }n\quad\text{and}\,\,		 
		\sup_{\epsilon\ge\epsilon_{n}}\frac{\sup_{\theta\in\Theta:\epsilon<d(\theta,\theta_0)\le 2\epsilon}\log\frac{\pi_n(\theta)}{\pi_n(\theta_0)}}{n\epsilon^{2}}\rightarrow 0.
	\end{equation}
\end{thm}

\proof See Appendix \ref{sec_appendix:proof_mode}.

\begin{rem}[Discussion on the convergence rate]
	Theorem \ref{thm:mode} establishes a general (not necessarily $\sqrt{n}$) rate of convergence for the distribution of posterior modes.
	The first part of Equation \eqref{eq:mode_rate} is similar to previous conclusions for M-estimators (\cite{wellner2013weak}). It shows that the convergence rate for $\hat\theta_{\bm w}$ is driven by the growth of $\mathcal{R}_n(\mathcal{M}_\epsilon(\theta_0))$ around $\epsilon=0$, which is determined by the richness of the function class $\{\log p_{\theta}:\theta\in\Theta\}$ around $\theta=\theta_0$. For example, with monotone densities, $\epsilon_n=n^{-1/2}$ and with convex densities in $\mathbb{R}^d$, $\epsilon_n=n^{2/(d+4)}$ for $d\le 3$, $\epsilon_n=n^{1/4}/[\log(n)]^{1/2}$ for $d=4$ and $\epsilon_n=n^{1/d}$ for $d>4$. We refer interested readers to \cite{wellner2013weak,pollard1991asymptotics} for more examples. 
	The second part of \eqref{eq:mode_rate} is less common, and it says that the convergence rate will also be affected by prior $\pi_n(\theta)$. In particular, a sufficient prior mass has to be put around $\theta=\theta_0$, otherwise the convergence rate will be slowed. 
\end{rem}


{
\begin{rem}[Errors from deep learning approximation]
We note that all previous results refer to the actual posterior, not the Deep Bootstrap approximation. In other words,  we do not consider the estimation error in obtaining $\hat\theta_{\bm w}$ using the deep learning approximations. Theoretically, there always exists a sufficiently large neural network whose approximation error is sufficiently small (\cite{hornik1989multilayer,guhring2020error}). Thus, if we allow its size to grow at a proper rate, we might show existence of a sequence of networks whose mapping converges at a rate no slower than $\epsilon_n$. The actual estimation error of the trained neural network, however, would need to incorporate the actual optimization of the network.
\end{rem}
}

\section{Deep Bootstrap in Bayesian Practice}
This section presents several stereotypical toy examples of inference about parameters determined by loss-functions. 
We aim to illustrate the potential of the deep bootstrap sampler for Bayesian inference.

\subsection{Bayesian Support Vector Machines}\label{sec:svm}
We first demonstrate the performance of the Deep Bootstrap sampler for Bayesian non-parametric learning (Section \ref{sec:nbl}) in binary classification tasks. 
Given data $\{(y_i,\bm x_i)\in\{-1,1\}\bigotimes\mathbb{R}^p\}_{i=1}^n$ where $\x_i$ denotes the covariates of the $i^{th}$ observation with a binary label $y_i\in\{-1,+1\}$, binary classification aims to predict $y$ when given $\x$ using the sign of $f(x)$,  where $f:\mathbb{R}^p\rightarrow \mathbb{R}$ is a function to be learned. Various loss functions have been designed to learn $f$, including the Support Vector Machine (SVM) loss (\cite{cortes1995support})
\begin{equation*}
L(y,f(\x))=\max\{0,1-yf(\x)\},
\end{equation*}
and the logistic loss (\cite{pearl1920rate})
\begin{equation*}
L(y,f(\x))=\log(1+e^{-yf(\x)}).
\end{equation*}
Suggested by \cite{rosasco2004loss}, we choose the SVM loss with a linear $f$,
i.e., we minimize the empirical loss $\sum_{i=1}^n l(\beta,\bm\theta;y_i,\x_i)$ with
\begin{equation}\label{eq:svm}
l(\beta,\bm\theta;y_i,\x_i)=\max\{0,1-y_i\left(\beta+\x_i'\bm\theta\right)\},
\end{equation}
where $\beta\in\mathbb{R}$ is the bias and $\bm\theta=(\theta_1,\cdots,\theta_p)'\in\R^p$ are the regression coefficients.
For the DP prior, following \cite{fong2019scalable}, we use the prior centering measure
\begin{align}
&F_\pi(y,\x)=F_\pi(\x)F_\pi(y),\quad\text{where}\label{eq:NPL_prior}
\\
&F_\pi(\x)=\frac{1}{n}\sum_{i=1}^n\delta_{\x_i},\quad\text{with }\delta_{\x}\text{  the Dirac delta measure centered at }\x,\nonumber
\\
&F_\pi(y)=\text{Bernoulli}(0.5).\nonumber
\end{align}
As discussed in \cite{fong2019scalable}, this choice of $F_\pi$ assumes that  $y,\x$ are independent and thus is equivalent to assuming $\bm\theta=\bm 0_p$ a priori, which  induces similar effects as shrinking priors on $\bm\theta$ (for example, $\|\bm\theta\|_1$ or $\|\bm\theta\|_2$). Regarding the choice of the concentration parameter $\alpha$, larger $\alpha$ represents stronger beliefs in the prior. Here, following \cite{fong2019scalable}, we set $\alpha=1.0$.

To generate simulated data sets, we adopt the setting in \cite{lyddon2019general} and extend to multivariate $\x$'s: we sample $n$ i.i.d. data points $\left\{(y_i, \bm x_i)\right\}_{i=1}^n$ from
\begin{equation}\label{eq:svm_generation}
\P\left(y_i=1\right)=\P\left(y_i=-1\right)=1/2, \quad \bm x_i\mid y_i\sim N(y_i\bm 1_p,\Sigma),
\end{equation}
where $\Sigma$ is a $p\times p$ matrix with all $1$'s on the diagonal and $\rho$'s off-diagonally. We consider $\rho=0$ (independent covariates) and a more challenging case $\rho=0.6$ (equi-correlated covariates). For brevity, results for the independent case is deferred to the Appendix (Section \ref{sec_appendix:svm}). 
Note that {with Equation \eqref{eq:svm_generation}, the inferential target $\theta_0=(\beta,\bm\theta)$ is a solution to the optimization problem defined in Equation \eqref{eq:theta0}, which does not have a closed form solution for the loss \eqref{eq:svm} and which does not necessarily satisfy $\beta=0,\bm\theta=\bm 1_p$. For example, when $p=10,\rho=0.6$, we have $\beta\approx 0,\bm\theta\approx0.2\times\bm 1_p$ by solving \eqref{eq:theta0} numerically. The misalignment between the inferential target $\theta_0=(\beta,\bm\theta)$ and the truth $(0,\bm 1_p)$ is not harmful for binary classification tasks as prediction is usually of more interest.
} 
We consider varying samples sizes $n\in\{50,500, 1000, 2000, 5000\}$ and dimensions of covariates $p\in\{10, 50, 100, 200, 500\}$.

\begin{figure}[t]
	\centering
	\includegraphics[width=.45\linewidth, height=.15\textheight]{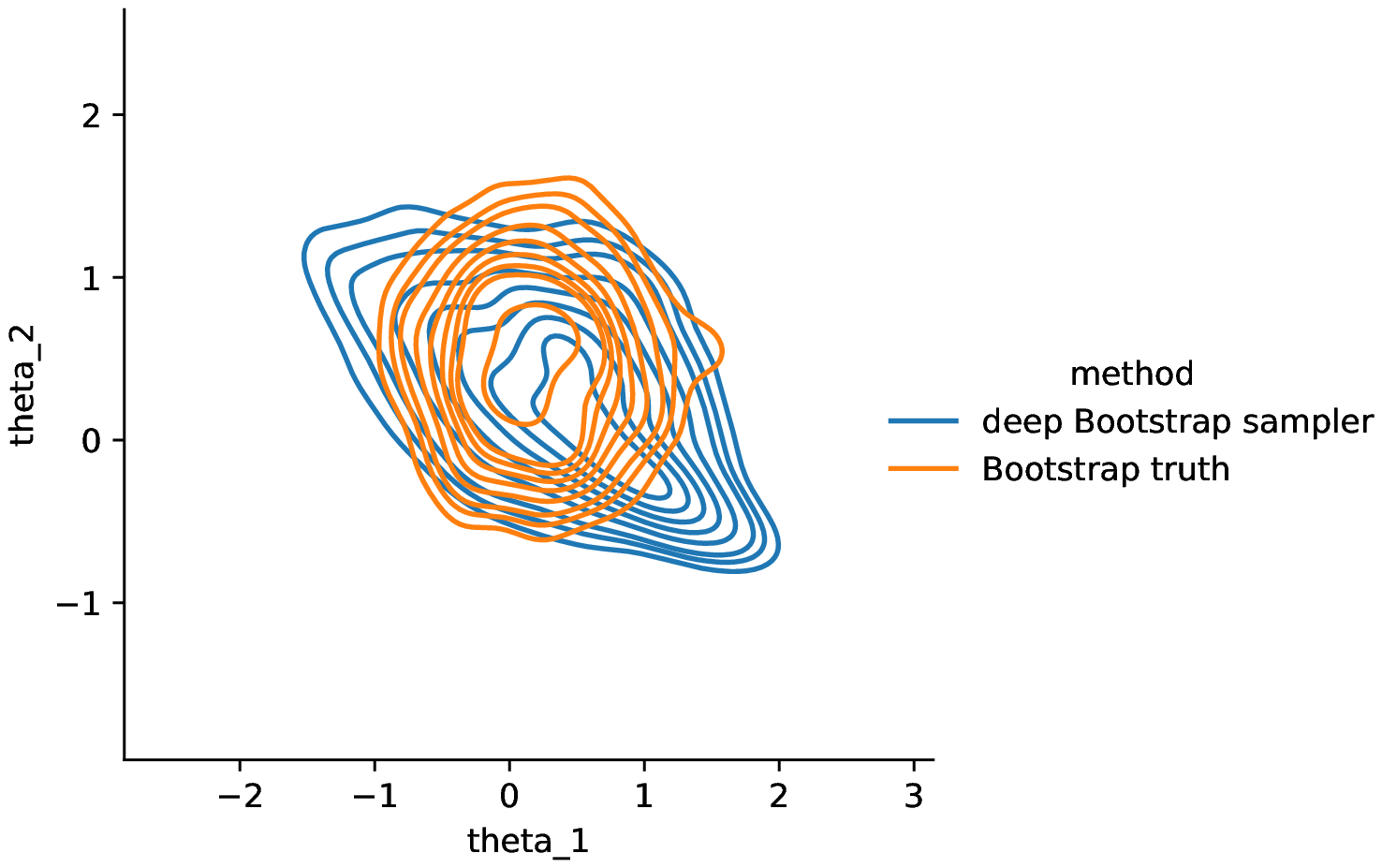}
	\includegraphics[width=.45\linewidth, height=.15\textheight]{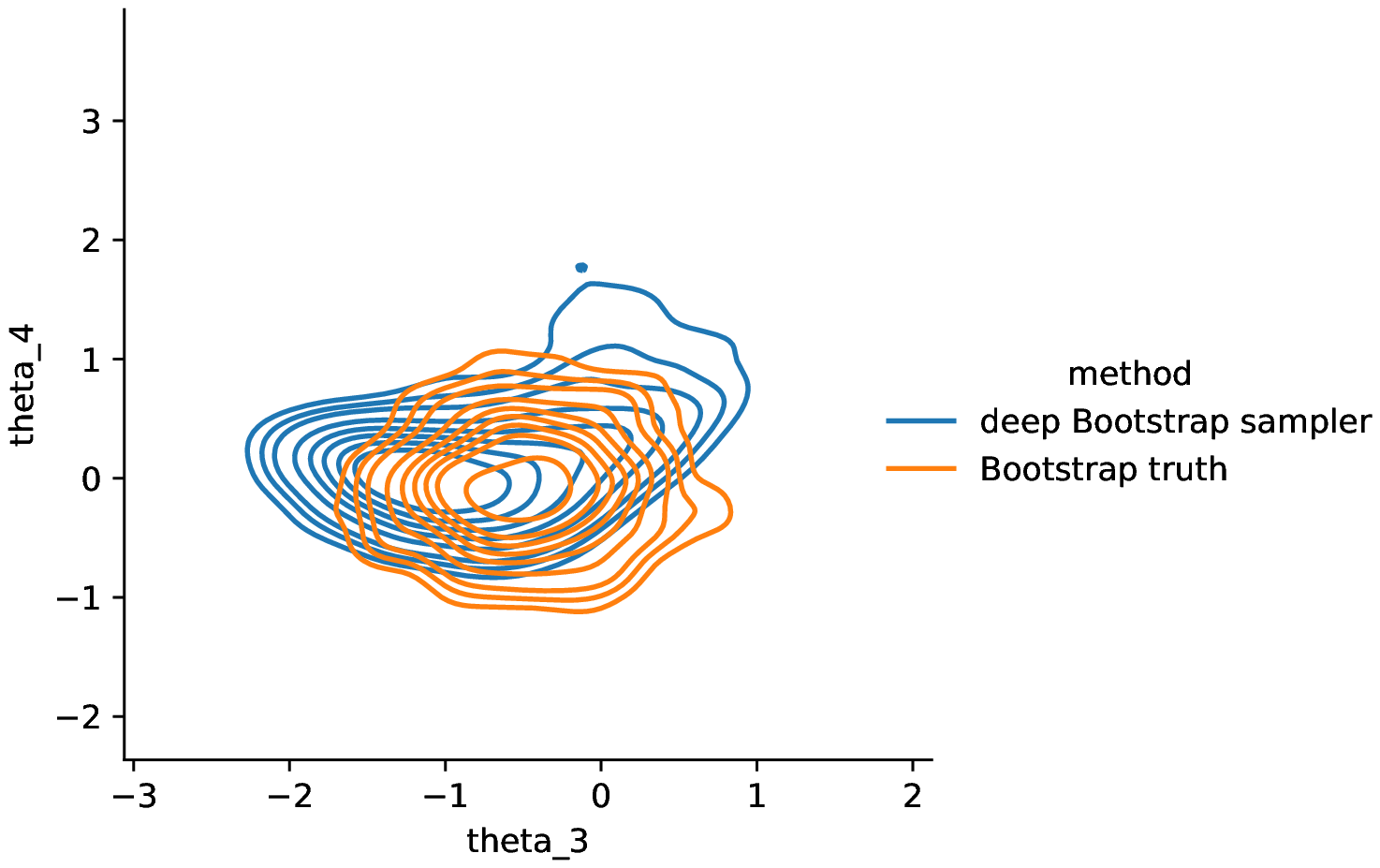}
	\caption{\small Two-dimensional posterior density plot for $\theta_i$'s from deep Bootstrap sampler and the truth for the Bayesian support vector machine example. We set $n=50,p=10,\rho=0.6,\alpha=1.0$. 
	}
	\label{fig:NPL_2d_density}
\end{figure}

We use the deep learning architecture (Figure \ref{fig:architecture}) introduced in \cite{shin2020scalable} to fit the Deep Bootstrap sampler. As discussed in \cite{shin2020scalable}, the benefit of this structure is that the re-introduction of weights at each hidden layer helps ``gradient flow in training deep neural networks'' and thus alleviates  any potential variance underestimation issues. Using the notation in Section \ref{sec:deep_bootstrap_sampling}, we set  $L-1=3$ fully connected (linear function + ReLU) hidden layers, each containing 128 neurons. Our sensitivity analysis in Appendix \ref{sec_appendix:complexity} shows that the network architecture has an impact on the approximating performance of the Deep Bootstrap sampler, yet such impact is minimal once the network complexity is moderately large (for example, $L-1=3,n_l=128,l=1,2,3$ sufficies for all experiments we tried). The implementation follows Algorithm \ref{alg:nonparam} and is coded via an optimized machine learning framework \texttt{PyTorch} (\cite{paszke2017automatic}). 
As suggested by \cite{shin2020scalable}, we use subgroup bootstrap with $S=100$ and $S'=10$. The subgroup bootstrap strategy significantly boosts the computational benefit of deep Bootstrap samplers (DBS), and does not hurt much its performance in our experiments. 
We set the number of Monte Carlo samples $K=100$, learning rate $\eta$ initialized at $0.0003$ and following a decay rate of $t^{-0.3}$ where $t$ is the current number of epoch (\cite{shin2020scalable}). In all $(n,p)$'s we tried, the training usually stabilizes after around 2\,000 iterations, and we set $T=4\,000$ to ensure convergence. 
RMSprop algorithm (\cite{graves2013generating}) is used to update the gradients instead of the vanilla gradient descent in Algorithm \ref{alg:nonparam}.

Sampled points $\{\theta^1,\cdots,\theta^N\}$ from the true bootstrap distribution  are generated following Algorithm 2 in \cite{fong2019scalable}. It requires solving an optimization problem
\begin{equation}\label{eq:svm_target}
\theta^j=\arg\max_{(\beta,\bm\theta)} \sum_{i=1}^n w_i\times l(\beta,\bm\theta;y_i,\x_i)
+\sum_{i=1}^{n'} w_{i+n}\times l(\beta,\bm\theta;y_i',\x_i'),\quad\forall j=1,2,\cdots,N,
\end{equation}
with $l(\beta,\bm\theta;y_i,\x_i)$ defined in \eqref{eq:svm}, $(\x_i',y_i')\stackrel{\text{iid}}{\sim}F_\pi(y,\x)$ where $F_\pi$ is defined in \eqref{eq:NPL_prior}, and the set of weights $(w_1,\cdots,w_{n+n'})$ are drawn from $Dir(1,\cdots,1,\alpha/n',\cdots,\alpha/n')$. Here $n'$ is the number of pseudo-samples and we set $n'=n$.
We solve the optimizatin problem \eqref{eq:svm_target} using the function \texttt{linear$\_$model.SGDClassifier} in Python library \texttt{sklearn} (\cite{scikit-learn}). This function allows various loss functions and different weight $w_i$'s for each sample point, and optimizes the loss function via stochastic gradient descent. We use the adaptive learning rate schedule (\cite{scikit-learn}) and tune the initial learning rate among $\{0.0001, 0.001, 0.01, 0.1\}$. The maximum number of epochs is set to $20\,000$ with early stopping turned on for saving computations.

To evaluate the performance of the Deep Bootstrap sampler, we first consider the two-dimensional density plots of $\theta_j$'s. Figure \ref{fig:NPL_2d_density} depicts an example where $n=50,p=10,\rho=0.6$. We observe that the Deep Bootstrap sampler captures the bootstrap posterior mean, but less so its variance. 
This issue has been discussed in \cite{shin2020scalable}. \cite{shin2020scalable} believe that it is caused by the vanilla feed-forward neural network which prevents variation in the input weights to {properly} transmit to the output layers as the neural network grows deeper, and they propose the network structure in Figure \ref{fig:architecture} to alleviate it. Here, we observe that this issue still persists when applied to Bayesian models. The study in Appendix \ref{sec_appendix:regularization} shows that for Bayesian models,  {as the sparsity-inducing prior grows stronger, the Deep Bootstrap sampler goes from accurate or slight over-estimation of the variance to more and more severe underestimation. }
We emphasize that, usually, the variance {mismatch} issue is not severe as long as  the regularization strength is properly selected (e.g., using the  BIC criterion (\cite{schwarz1978estimating}) as in the next example). 
In this example (with $\alpha=1.0$), it does not affect the predictive inference on testing data, which is shown in Table \ref{table:svm}.

\begin{table}[t]
	\centering
	\small
	\scalebox{0.7}{
		\begin{tabular}{l|l|l|l|l|l|l|l|l} 
			\hline\hline
			\bf \large Setting&&\multicolumn{7}{|l}{  \cellcolor[gray]{0.8}\large \bf Equi-correlated $\rho=0.6$}    \\
			\cline{1-9}
			& metric   
			& accuracy & precision & recall & F1 & ROC & PR & time\\ 
			\hline
			\multirow{2}{*}{$p=10,n=50$}   & DBS                     
			& 0.83 & 0.82 & 0.86 & 0.83 & 0.91 & 0.92  &48.59+0.58\\
			& WLB                 
			& 0.86 & 0.85 & 0.90 & 0.87 & 0.94 & 0.94 &110.99\\ 
			\hline
			\multirow{2}{*}{$p=50,n=500$}  & DBS       
			&  0.87     &    0.88 & 0.87 & 0.88 & 0.94 & 0.94       &      97.01+0.65       \\
			& WLB         
			&   0.86   &   0.87 & 0.87 & 0.87 & 0.94 & 0.94     &      208.10             \\ 
			\hline
			\multirow{2}{*}{$p=100,n=1000$}  & DBS                
			& 0.86& 0.88 & 0.85 & 0.86 & 0.93 & 0.94  &194.80+0.64\\
			& WLB                
			& 0.85 & 0.83 & 0.85 & 0.86 & 0.93&0.94 & 436.86 \\ 
			\hline
			\multirow{2}{*}{$p=200,n=2000$} & DBS        
			& 0.87 & 0.87 & 0.87 & 0.87 & 0.93 & 0.93   &  315.19+0.66\\
			& WLB              
			& 0.86 & 0.86 & 0.86 & 0.86 & 0.93 & 0.93   &  1187.21 \\
			\hline
			\multirow{2}{*}{$p=500,n=5000$} & DBS      
			& 0.89 & 0.89 & 0.89 & 0.89 & 0.95 & 0.92   &  565.05+0.63\\
			& WLB          
			& 0.87 & 0.86 & 0.87 & 0.87 & 0.94 & 0.93   &  5677.71 \\
			\hline\hline
		\end{tabular}
	}
	\caption{\small Evaluation of approximation properties based on 10 independent runs for Bayesian support vector machine example. DBS standards for `deep Bootstrap sampler'.  WLB standards for samples from the true bootstrap distribution. `Bias'   refers to the $l_1$ distance of estimated posterior means. `ROC' refers to the area under the receiver operating characteristic curve;  `PR' refers to the area under the precision-recall curve. The last column in each setting represents the time (in seconds) to generate 10\,000 sample points. For deep Bootstrap sampler, time reported is in the form of training time $+$ sampling time.}
	\label{table:svm}
\end{table}

In Table \ref{table:svm}, we measure the ability of the Deep Bootstrap sampler to approximate the bootstrap target in terms of predictive performance. {For each $x_i$, we assign an `average' label by the majority vote based on samples of $(\beta,\bm\theta)$.} We report quantitative metrics that reflect the quality of this average (accuracy, precision, recall, F1 score). 
{In addition, we report metrics} that reflect the shape of the whole Deep Bootstrap  posterior (the area under the receiver operating characteristic curve and the precision-recall curve). The metrics are calculated for different choice of $n,p$'s, and on a separate testing set  of size 100. The results are summarized in Table \ref{table:svm} which shows  that samples generated from the Deep Bootstrap sampler are of high quality for various prediction-related metrics and  are close to the target. In terms of   predictive performance,   the actual bootstrap and DBS are comparable. However, the computing times are dramatically different. We observe that the timing benefit of DBS increases as $n$ or $p$ grow larger. 

In summary, this example shows that DBS approximates  the true bootstrap posterior well in terms of predictive performance. However, the deep sampler is dramatically faster,  especially in large $(n,p)$ settings. The variance {mismatch} 
issue mentioned in \cite{shin2020scalable} exists, but is not fatal.

\subsection{Bayesian LAD Regression}\label{sec:LAD}
Least squares regression estimators  tend  to be less robust to outliers or heavy-tailed errors. When robustness becomes a concern, M-estimators (\cite{huber2009robust}) are often used instead of least-square estimators. Given data $\{(y_i,\bm x_i)\in\R\bigotimes\mathbb{R}^p\}_{i=1}^n$ where $\x_i$ denotes the covariates of an observation with response $y_i$, a regression M-estimator is the minimizer of the loss
\begin{equation}
l(\beta,\bm\theta;y_i,\x_i)=g(y_i-\beta-\x_i^\top\bm\theta).
\end{equation}	
where $\beta\in\R$ is the intercept, $\bm\theta=(\theta_1,\cdots,\theta_p)\in\R^p$ is the regression coefficient vector, and  $g: \mathbb{R}\rightarrow [0, \infty)$ is a residual function that satisfies (i) $g(0)=0$, (ii) $g(t)=g(-t),\forall t\in\R$, (iii) $g$ is monotonically increasing (\cite{rousseeuw2005robust}). 
Statistical literature on robustness has proposed a variety of residual functions $g$, including the Huber function (\cite{huber1964robust})
\begin{equation*}
g(t;\delta)=
\begin{cases}
\frac{1}{2}t^2,\quad&\text{for }|t|\le\delta\\
\delta\left(|t|-\frac{1}{2}\delta\right),\quad&\text{otherwise}
\end{cases}
\end{equation*}
and the absolute value function (\cite{boscovich1757litteraria})
$
g(t)=|t|
$. Here, we consider the case where $g$ is the absolute value function, i.e.,  least absolute deviation (LAD) regression (\cite{boscovich1757litteraria}). A penalized LAD regression model is investigated in \cite{wang2006regularized,zou2006adaptive,wang2007robust,lambert2011robust,wang2013l1}, which minimizes
\begin{equation}\label{eq:penalized_LAD}
\sum_{i=1}^n\left|y_i-\beta-\bm\theta^\top\bm x_i\right|+\lambda\sum_{j=1}^p|\theta_j|.
\end{equation}
Note that the intercept $\beta$ is excluded from the penalty term as suggested by \cite{wang2006regularized}. In \eqref{eq:penalized_LAD}, the regularization strength $\lambda$ can be selected from the classical BIC criteria as recommended by \cite{schwarz1978estimating,lambert2011robust}.

Viewing \eqref{eq:penalized_LAD} from a Bayesian viewpoint, minimizing \eqref{eq:penalized_LAD} is equivalent to estimating the mode of a Bayesian model with a loss
\begin{equation}\label{eq:gibbs_loss}
l(\beta,\theta;y_i)=\left|y_i-\beta-\bm\theta^\top\bm x_i\right|,
\end{equation}
and a prior 
\begin{equation}\label{eq:gibbs_prior}
\pi(\bm\theta)=\prod_{j=1}^p\left[\frac{\lambda}{2}e^{-\lambda|\theta_j|}\right].
\end{equation}
However, one might be interested in not only the posterior mode but also uncertainty quantification. The Gibbs posterior defined by \eqref{eq:gibbs} in Section \ref{sec:loss} provides one such uncertainty measurement. We compute the Gibbs posterior using MCMC. Another possibility is to obtain approximations, either by directly solving the optimization problem defined in \eqref{eq:theta_hat}, or by using the deep Bootstrap sampler (Algorithm \ref{alg:nonparam}). This examples investigates these three objects for uncertainty quantification.

We simulate data following the settings in \cite{lambert2011robust}: $n$ i.i.d. data points $\{(y_i,\x_i)\}_{i=1}^n$ are generated from 
\begin{equation}
y_i=\beta^*+\x_i^\top\bm\theta^*+\sigma\epsilon_i,\quad
\x_i\stackrel{\text{iid}}{\sim}N(0,\Sigma),
\end{equation}
where $\beta^*=1$, $\bm\theta^*=(1.5, 2, 3,0,0,\cdots,0)\in\R^p$, $\Sigma$ is a Toeplitz matrix with $\Sigma_{i,j}=(0.5)^{|i-j|}$. 
Following \cite{lambert2011robust}), we consider two challenging cases with heavy-tailed noise distribution:
\begin{itemize}
	\item Model 1: large outliers. $\epsilon_i=v_i/\sqrt{\text{var}(v_i)}$, $\sigma=9.67$, and $v_i\stackrel{\text{iid}}{\sim}0.9N(0,1)+0.1N(225)$.
	\item Model 2: sensible outliers. $\epsilon_i=v_i/\sqrt{\text{var}(v_i)}$, $\sigma=9.67$, and $v_i\stackrel{\text{iid}}{\sim}Laplace(1)$.
\end{itemize}
One may show that the solution to \eqref{eq:theta0} equals the truth (i.e., $\theta_0=(\beta,\bm\theta)=(\beta^*,\bm\theta^*)$) under some mild conditions as in \cite{pollard1991asymptotics,gross1979least,ruzinsky1989strong}. We note that this equality holds for both Model 1 and 2. 
We investigate different choices of $n\in\{8, 10, 20, 50\}$ and $p\in\{100, 200, 500, 1000\}$. 

Both the Gibbs posterior \eqref{eq:gibbs} and bootstrap samples \eqref{eq:theta_hat}  use the loss \eqref{eq:gibbs_loss} and the prior \eqref{eq:gibbs_prior}, where the regularization strength $\lambda$ in prior \eqref{eq:gibbs_prior} is set to the one that minimizes the BIC criteria (\cite{schwarz1978estimating}) among $\log(\lambda)\in\{-6,\cdots,1\}$ (an equi-spaced sequence starting at -6, ending at 1 and of length 20), as suggested by \cite{lambert2011robust}.
For the Gibbs posterior, we set the learning rate $\alpha=1$ to match \eqref{eq:penalized_LAD} which is used to generate bootstrap samples. We refer readers to \cite{bissiri2016general} and \cite{lyddon2019general} for more discussion on the problem of determining $\alpha$.

In practice, the Gibbs posterior is generated using the Metropolis-Hasting algorithm described in \cite{chernozhukov2003mcmc}. We implement it in Python. For faster convergence, the starting value for $\theta_0,\bm\theta$ are set to the mode of the Gibbs posterior  solved by \texttt{QuantileRegressor}. The proposal density $q(x\mid y)$ is set to
the density of $N(y,\sigma'I_p)$. We set $\sigma'=0.1$ for small data sets and $\sigma'=0.01$ for large ones as the true Gibbs posterior tends to have smaller variance as the dimension grows. 
We run MCMC for $1\,000\,000$ iterations and discard the first $10\,000$ as burn-in. Gelman-Rubin diagnostic (\cite{gelman1992inference}) computed by R package \texttt{coda} (\cite{coda}) is checked to ensure convergence.

\begin{figure}[t]
	\centering
	\includegraphics[width=.45\linewidth, height=.15\textheight]{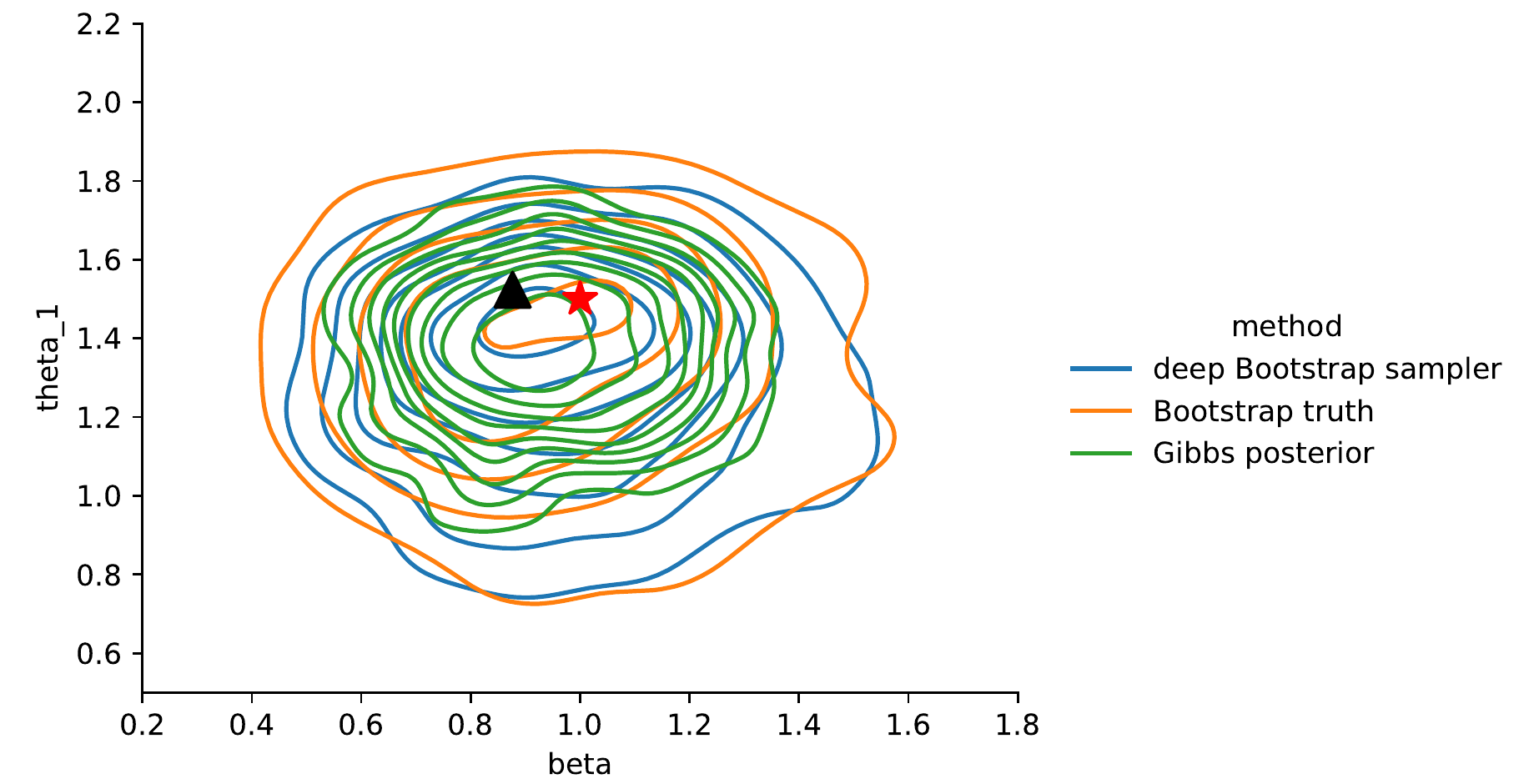}
	\includegraphics[width=.45\linewidth, height=.15\textheight]{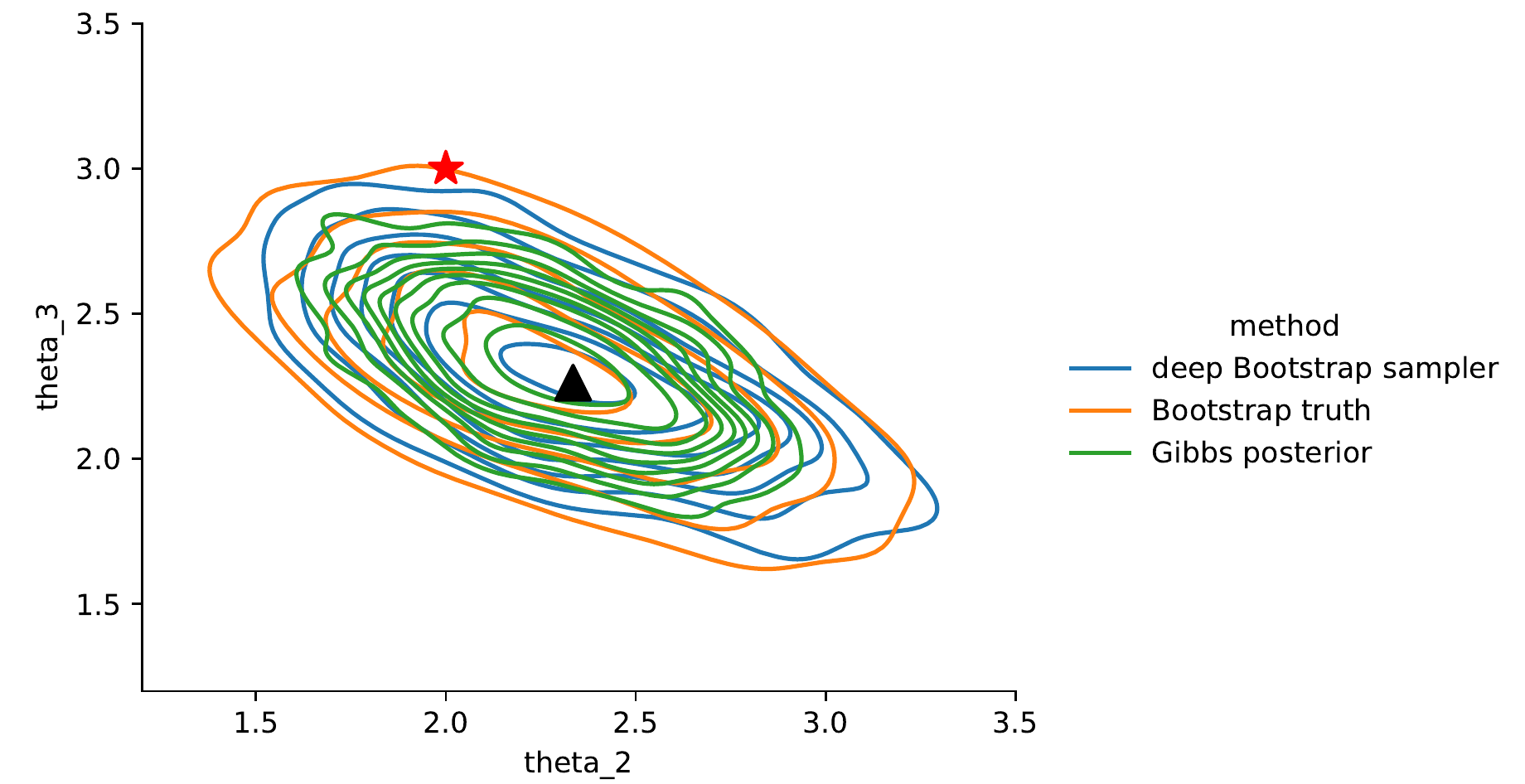}
	\caption{\small Two-dimensional posterior density plot for $\theta_i$'s in Model 1 for Bayesian LAD example. We set $n=100,p=8$. The location of true parameters are marked in red star. {The location of the minimizer of Equation \eqref {eq:penalized_LAD} are marked in black triangle.}}	
	\label{fig:LAD_2d_density}
\end{figure}

For the Bootstrap samples, we solve the optimization problem \eqref{eq:theta_hat} with $w\sim n\times Dir(1,1,\cdots,1)$, loss \eqref{eq:gibbs_loss}, and prior \eqref{eq:gibbs_prior} using the built-in function \texttt{QuantileRegressor}. This function solves \eqref{eq:theta_hat} by linear programming, and we use its default interior-point solver with default parameters.
For the Deep Bootstrap sampler, we implement Algorithm \ref{alg:nonparam} using \texttt{Pytorch}, with a subgroup size $S=100$. The other settings, including network architecture, training schedule and hyper-parameters, are identical to the Bayesian support vector machine example. 

Let us first consider the posterior densty of $\theta_j$'s. Figure \ref{fig:LAD_2d_density} shows the two-dimensional posterior density plot in Model 1. Results for Model 2 are very similar and are included in the Appendix \ref{sec_appendix:LAD}. 
We observe that samples from all three methods are centered around the same location. {The center is close to the minimizer of Equation \eqref {eq:penalized_LAD}, which is different from the truth due to shrinkage effects introduced from the prior.}
The Gibbs posterior tends to have the smallest variance, in contrast to the true Bootstrap samples which has the largest variance.  


To quantitatively compare each method, we report the bias, lengths of 90\% credible intervals and their coverage for various choice of $(n,p)$'s in Table \ref{table:LAD}. We separate between active and inactive coordinates. Table \ref{table:LAD} also includes the time cost for each method. For fairness, we compare the time required to generate the same number of effective samples. Since each sample from boostrap is independent, we treat  the total number of samples as the effective sample size for both bootstrap samplers. The effective sample size for Gibbs posterior is calculated using R package \texttt{coda}.
Table \ref{table:LAD} shows the deep Bootstrap sampler is much faster than both the true bootstrap and the Gibbs posterior, especially in large datasets. 

We conclude that, in this example, DBS reconstructions could be a viable alternative to Gibbs posteriors. The Deep Bootstrap sampler achieves a similar bias and a larger variance, but is much faster than the Metropolis-Hastings algorithm.


\begin{table}[t]
	\centering
	\small
	\scalebox{0.65}{
		\begin{tabular}{l|l|l|l|l|l|l|l|l||l|l|l|l|l|l|l} 
			\hline\hline
			\bf \large Setting&&\multicolumn{7}{l||}{  \cellcolor[gray]{0.8}\large \bf Model 1}   &\multicolumn{7}{|l}{  \large\cellcolor[gray]{0.8} \bf Model 2}   \\
			\cline{1-16}
			\multirow{2}{*}{}      & \multirow{2}{*}{metric}   & \multicolumn{2}{l|}{coverage} & \multicolumn{2}{l|}{length of 90\% CI} & \multicolumn{2}{l|}{bias}  & time
			& \multicolumn{2}{l|}{coverage} & \multicolumn{2}{l|}{length of 90\% CI} & \multicolumn{2}{l|}{bias} & time\\ 
			\cline{3-16}
			&                  & $+$     & $-$                     & $+$     & $-$                              & $+$     & $-$  &
			& $+$     & $-$                     & $+$     & $-$                              & $+$     & $-$      &            \\ 
			\hline
			\multirow{2}{*}{$p=8,n=100$}   & DBS                     & 0.90 &      0.98            &1.10  &    1.23                     & 0.23 &    0.268  & 50.51+0.63
			& 0.90& 0.88&3.43 & 3.82&0.79 &  0.92   & 57.20+0.76 \\
			& WLB         & 0.98 &      1.00      & 1.37 &          1.52                 &  0.23&         0.26    &534.08
			&0.90 &0.95 & 4.19& 4.71&0.79 & 0.90 & 546.73\\ 
			& Gibbs                  &  0.78&    0.80      &  0.71&    0.80             & 0.23&    0.29     & 42.71
			& 0.45&0.43&1.22 &1.32 & 0.82&0.96  & 82.12\\ 
			\hline
			\multirow{2}{*}{$p=10,n=100$}  & DBS         & 0.85&0.85 & 0.99&1.15 &0.26 &   0.33   &       79.83+1.01
			&   0.90    &         0.83              &  3.20     &     3.93                           &    0.77  & 1.14      &   68.39+0.85         \\
			&WLB        & 0.95& 0.97& 1.27& 1.44& 0.25&       0.34    &889.63
			&  0.95     &       0.93                &   4.11   &     4.86&         0.74         &  1.18  &   780.12    \\ 
			& Gibbs                  & 0.75 &    0.65    &0.67  &     0.76                & 0.27 &     0.34     & 58.58
			& 0.43&0.33 &1.14 & 1.39& 0.79& 1.16 & 103.04\\ 
			\hline
			\multirow{2}{*}{$p=20,n=200$}  & DBS         & 0.90&0.88 & 0.72&0.76 &0.18 &   0.19   &    70.05+0.86   
			&   0.78   &        0.82          &  2.28     &     2.46                         &    0.66  & 0.71      &   74.97+0.94         \\
			& WLB       & 1.00& 0.95& 0.93& 1.00& 0.19&      0.20    &2891.89
			&  0.93    &      0.91           &   2.94   &     3.29&        0.69      &   0.73   &    3041.20    \\ 
			& Gibbs                  & 0.35 &  0.35     & 0.26 &   0.27           & 0.19&    0.29      &696.23
			& 0.43& 0.39& 0.85& 0.93& 0.64&  0.71&1550.77 \\ 
			\hline
			\multirow{2}{*}{$p=50,n=500$}  & DBS                      &   0.78  &      0.83          &  0.41   &      0.43    &  0.13    &         0.12 & 74.43+0.91
			& 0.85& 0.82& 1.35&1.42 & 0.42&    0.41  & 81.64+1.01\\
			& WLB     & 0.93     &      0.96          &  0.60   &      0.63                 &  0.13   &        0.12&> 6 hours
			& 0.95&0.95 & 1.89& 2.04& 0.42&  0.42   &> 6 hours   \\ 
			& Gibbs                  & 0.68&   0.74   & 0.32 &    0.34               & 0.13&    0.12      &703.52
			&0.38 & 0.43& 0.54& 0.59& 0.40&  0.41& 1529.81\\ 
			\hline
			\multirow{2}{*}{$p=50,n=1000$} & DBS                      &   0.93   &        0.80            &  0.27  &     0.30             &   0.08    &            0.09 & 75.08+0.87
			&0.88 & 0.86& 0.86& 0.95& 0.23&  0.25  &  81.90+0.93\\
			& WLB     &   0.98    &            0.91          &   0.38  &        0.41        &   0.15 &        0.20 & > 2 days
			& 0.98& 0.94&1.17&1.24 &0.23 &     0.26   &  > 2 days \\
			& Gibbs                  &  0.73&    0.68      & 0.21&         0.23                & 0.08 &   0.09        & 545.54
			&0.38 &0.47 &0.35 & 0.39& 0.24& 0.25 & 1140.76\\ 
			\hline\hline
		\end{tabular}
	}
	\caption{\small Evaluation of approximation properties in different settings based on 10 independent runs of  Bayesian LAD regression. DBS standards for `Deep Bootstrap Sampler'. WLB standards for samples from the true bootstrap distribution. Coverage stands for the empirical coverage of 90\% credible intervals. `Bias'   refers to the $l_1$ distance between estimated posterior means and the truth. We denote with $+$ an average over active coordinates, and with $-$ an average over inactive coordinates. Times reported are the number of seconds (unless otherwise noted) taken to generate 10\,000 effective samples for each procedure. For deep Bootstrap sampler, time reported is in the form of training time $+$ sampling time. 
	}
	\label{table:LAD}
\end{table}

\section{Discussion}
This paper surveys several recent contributions to the Bayesian literature on learning about parameters defined by loss functions. We highlighted a new promising direction for Bayesian computation using generative bootstrap.
We demonstrated the potential of this new strategy on several examples. This paper aims to draw practitioners' attention towards  posterior sampling techniques beyond the traditional MCMC technology.

\bibliographystyle{apa}
\bibliography{references}

\newpage

\appendix

\section{Proofs}

\subsection{Notations}
First let us recall some notations. For a random variable $X$ on a probability space $(\Omega,\Sigma,P)$, we denote $\mathbb{P}(A)=\int_A dP$ for any $A\in\Sigma$, and $Ph=\int_\Omega hdP$ for any $h:\Omega\rightarrow\R$. We denote $[n]=\{1,2,\cdots,n\}$. We use $\|\cdot\|$ to denote Euclidean norm.

\subsection{Proof of Lemma \ref{lemma:unique_minimizer}}
\label{sec_appendix:unique_minimizer_proof}
Recall that
$$
p_{\theta}^{(n),\bm w}(\Xn)=\prod_{i=1}^np_{\theta}^{w_i}(X_i)\quad\text{and}\quad
p_{\theta_0}^{(n),1-\bm w}(\Xn)=\prod_{i=1}^np_{\theta_0}^{1-w_i}(X_i).
$$
By definition, we have
\begin{align}
C_{\theta,\bm w}
&=P_{\theta}^{(n)}\frac{p_{\theta_{0}}^{(n),1-\bm w}(\Xn)}{p_{\theta}^{(n),1-\bm w}(\Xn)}=\int_{\mathcal X} p_{\theta_{0}}^{(n),1-\bm w}(\Xn)p_{\theta}^{(n),\bm w}(\Xn)d\Xn
\nonumber
\\&=
P_{\theta_{0}}^{(n)}\frac{p_{\theta}^{(n),\bm w}(\Xn)}{p_{\theta_{0}}^{(n),\bm w}(\Xn)}
=P_{\theta_{0}}^{(n)}\left[\prod_{i=1}^{n}\frac{p_{\theta}^{w_i}(X_i)}{p_{\theta_{0}}^{w_i}(X_i)}\right]
\stackrel{(a)}{=}\prod_{i=1}^{n}\left[P_{\theta_{0}}\left(\frac{p_{\theta}^{w_i}(X_i)}{p_{\theta_{0}}^{w_i}(X_i)}\right)\right],
\label{eq_appendix:C_theta}
\end{align}
where $(a)$ follows from the independence of $X_i$'s. For any fixed $\bm w$, recall that $\theta^*_{\w}$ defined in \eqref{eq:dfn_theta_W^*} is the minimizer of the function $f_{\bm w}(\cdot)$, where
\begin{equation*}
f_{\boldsymbol{w}}(\theta)  =-P_{\theta_{0}}^{(n)}\log\frac{\tilde{p}_{\theta}^{(n),\boldsymbol{w}}(\Xn)}{p_{\theta_{0}}^{(n)}(\Xn)}.
\end{equation*}
Notice that
\begin{align*}
f_{\boldsymbol{w}}(\theta) =-P_{\theta_{0}}^{(n)}\log\frac{\tilde{p}_{\theta}^{(n),\boldsymbol{w}}(\Xn)}{p_{\theta_{0}}^{(n)}(\Xn)}
&=-P_{\theta_{0}}^{(n)}\left[\sum_{i=1}^{n}-\log\frac{p_{\theta_{0}}^{w_i}(X_{i})}{p_{\theta}^{w_i}(X_{i})}-\log C_{\theta,\bm w}\right]\\
& \stackrel{(a)}{=}\sum_{i=1}^n\left[P_{\theta_{0}}\left(\log\frac{p_{\theta_{0}}^{w_i}(X_i)}{p_{\theta}^{w_i}(X_i)}\right)+\log P_{\theta_{0}}\left(\frac{p_{\theta}^{w_i}(X_i)}{p_{\theta_{0}}^{w_i}(X_i)}\right)\right]\\
& \stackrel{(b)}{\ge}\sum_{i=1}^n\left[P_{\theta_{0}}\left(\log\frac{p_{\theta_{0}}^{w_i}(X_i)}{p_{\theta}^{w_i}(X_i)}\right)+ P_{\theta_{0}}\left(\log\frac{p_{\theta}^{w_i}(X_i)}{p_{\theta_{0}}^{w_i}(X_i)}\right)\right]=0,
\end{align*}
where $(a)$ uses Equation \eqref{eq_appendix:C_theta}, and $(b)$ follows from Jensen's inequality. From properties of Jensen's inequality, $(b)$ achieves equality if and only if $\frac{p_{\theta}^{w_i}(X_i)}{p_{\theta_{0}}^{w_i}(X_i)}$ is degenerate for every $i\in[n]$, i.e., there exist constants $c_{w_i}$ for $i=1,2,\cdots,n$ such that
\[
\mathbb{P}_{\theta_0}\left(\frac{p_{\theta}^{w_i}(X_i)}{p_{\theta_{0}}^{w_i}(X_i)}=c_{w_i}\right)=1,\quad\forall i=1,2,\cdots,n.
\]
From assumption (1), there must exist $k\in[n]$ such that $w_k>0$ and thus,
$
\mathbb{P}_{\theta_0}\left(\frac{p_{\theta}(X_k)}{p_{\theta_0}(X_k)}=c_{w_k}^{1/w_k}\right)=1$,
i.e. $\frac{p_{\theta}(X_k)}{p_{\theta_0}(X_k)}$ is degenerate. From assumption (2), we then conclude that when $(b)$ achieves equality, we must have  $\theta=\theta_0$. Notice that setting $\theta=\theta_0$ indeed achieves equality in $(b)$. 
Thus, $f_{\boldsymbol{w}}(\theta)$ is minimized iff $\theta=\theta_0$. This means that $\theta_0$ is the unique minimizer of $f_{\boldsymbol{w}}(\theta) $.

\subsection{Proof of Theorem \ref{thm:concentration}}
\label{sec_appendix:concentration_proof}
First, we introduce some notation and Assumptions. For any $\bm w$, let us define the usual $\epsilon$-KL neighborhood around $\wt{P}_{\theta_0}^{(n),\w}$  as 
\[
B\left(\epsilon,\bm w,\wt{P}_{\theta_0}^{(n),\w},P_{\theta_{0}}^{(n)}\right)
=\left\{ \wt{P}_{\theta}^{(n),\bm w}\in\mathcal{\wt{P}}^{(n),\bm w}:K\left(\bm w,\theta_{0}\right)\le n\epsilon^{2},V\left(\bm w,\theta_{0}\right)\le n\epsilon^{2}\right\} 
\]
where 
\[
K\left(\bm w,\theta_{0}\right)=P_{\theta_{0}}^{(n)}\log\frac{\wt{p}_{\theta_0}^{(n),\w}}{\wt{p}_{\theta}^{(n),\w}},
\quad\text{and}\quad
V\left(\boldsymbol{w},\theta_{0}\right)=P_{\theta_{0}}^{(n)}\left|\log\frac{\wt{p}_{\theta_0}^{(n),\w}}{\wt{p}_{\theta}^{(n),\w}}-K\left(\boldsymbol{w},\theta_{0}\right)\right|^{2}.
\]
We denote with $Q_\theta^{(n)}$ a probability measure defined through $dQ_\theta^{(n),\bm w}=\frac{p_{\theta_0}^{(n)}}{\tilde{p}_{\theta_0}^{(n),\bm w}}dP_\theta^{(n)}$.

\begin{assumption}[Existence of suitable semi-metric]\label{assump_appendix:metric}
	There exists a semi-metric $d(\cdot,\cdot)$ and some constant $C>0$ such that 
	for any sufficiently large $n\in\mathbb{N}$ and
	any $\w\in R_{n}(c_0)$, 
	\[
	d\left(\wt{P}_{\theta}^{(n),\w},\wt{P}_{\theta_0}^{(n),\w}\right)\ge C\|\theta-\theta_0\|,\quad \theta\in\Theta.
	\]
\end{assumption}

\begin{assumption}[Existence of suitable tests]\label{assump_appendix:test}
	For any $\epsilon>\epsilon_{n}$, there
	exists a test $\phi_{n}$ (depending on $\epsilon$) such that for
	any $J\in\mathbb{N}_0$, 
	\[
	P_{\theta_{0}}^{(n)}\phi_{n}\lesssim e^{-n\epsilon^{2}/4}
	\quad\text{and}\quad
	\sup_{\w\in R_{n}(c_0)}\sup_{\wt{P}_{\theta}^{(n),\w}:\,d\left(\wt{P}_{\theta}^{(n),\w},\,\wt{P}_{\theta_0}^{(n),\w}\right)>J\epsilon}Q_{\theta}^{(n),\bm w}\left(1-\phi_{n}\right)\le e^{-nJ^{2}\epsilon^{2}/4}.
	\]
\end{assumption}

\begin{assumption}[Prior regularization]\label{assump_appendix:prior}
	There exists constant $L>0$ such that for
	any sufficiently large $n,j\in\mathbb{N}$, 
	\[
	\sup_{\boldsymbol{w}\in R_{n}(c_0)}\frac{\wt{\Pi}_{\bm w}\left(\theta\in\Theta:j\epsilon_{n}<d\left(\wt{P}_{\theta}^{(n),\bm w},\wt{P}_{\theta_0}^{(n),\bm w}\right)\le\left(j+1\right)\epsilon_{n}\right)}{\wt{\Pi}_{\bm w}\left(B\left(\epsilon,\boldsymbol{w},\wt{P}_{\theta_0}^{(n),\bm w},P_{\theta_{0}}^{(n)}\right)\right)}\le Le^{n\epsilon_{n}^{2}j^{2}/8}.
	\]
\end{assumption}

\begin{proof}
	
	Notice that
	\begin{align}
	&
	P_{\theta_{0}}^{(n)}P_{\bm w}\left[\Pi_{\boldsymbol{w}}\left(\theta\in\Theta:\,\,\|\theta-\theta_0\|\ge M_{n}\epsilon_{n}\mid\Xn\right)\right]
	\nonumber\\
	= & P_{\boldsymbol{w}}P_{\theta_{0}}^{(n)}\left[\Pi_{\boldsymbol{w}}\left(\theta\in\Theta:\,\,\|\theta-\theta_{0}\|\ge M_{n}\epsilon_{n}\mid\Xn\right)\mathbb{I}\left(\boldsymbol{w}\in R_{n}\right)\right]
	\\&+P_{\boldsymbol{w}}P_{\theta_{0}}^{(n)}\left[\Pi_{\boldsymbol{w}}\left(\theta\in\Theta:\,\,\|\theta-\theta_{0}\|\ge M_{n}\epsilon_{n}\mid\Xn\right)\mathbb{I}\left(\boldsymbol{w}\notin R_{n}\right)\right]
	\nonumber\\
	= & U_1+U_2,\label{eq_appendix:thm_together}
	\end{align}
	where 
	$
	U_1=P_{\boldsymbol{w}}P_{\theta_{0}}^{(n)}\left[\Pi_{\boldsymbol{w}}\left(\theta\in\Theta:\,\,\|\theta-\theta_{0}\|\ge M_{n}\epsilon_{n}\mid\Xn\right)\mathbb{I}\left(\boldsymbol{w}\in R_{n}\right)\right]$
	and
	$U_2=P_{\boldsymbol{w}}\left[\mathbb{I}\left(\boldsymbol{w}\notin R_{n}\right)\right].$  Now, we will bound  $U_1$ and $U_2$ separately.

	From Lemma \ref{lemma:unique_minimizer}, we have $\theta^*_{\bm w}=\theta_0$ for any $\bm w$. 
	Thus, from
	Assumption \ref{assump_appendix:test}, \ref{assump_appendix:prior} and Theorem 11.1 in \cite{kaji2021metropolis}, we have for every sequence of constants $M_n\rightarrow\infty$, as $n\rightarrow\infty$, 
	\begin{equation}\label{eq_appendix:posterior_concentration}
	P_{\theta_{0}}^{(n)}
	\left[\Pi_{\boldsymbol{w}}\left(\tilde{P}_{\theta}^{(n),\w}:\,\,d\left(\wt{P}_{\theta}^{(n),\w},\wt{P}_{\theta_0}^{(n),\w}\right)\ge M_{n}\epsilon_{n}\mid\Xn\right)
	\mathbb{I}\left(\bm w_n\in R_n(c_0)\right)\right]
	\rightarrow0,
	\end{equation}
	where $\Pi_{\bm w}(\cdot\mid X^{(n)})$ is the probability measure with a density
	\[
	\pi_{\bm w}\left(\theta\mid X^{(n)}\right)\propto\tilde p_\theta^{(n),\bm w}\left(X^{(n)}\right)\tilde\pi_{\bm w}(\theta).
	\]
	From \eqref{eq_appendix:posterior_concentration} and Assumption \ref{assump_appendix:metric}, we have 
	\begin{equation}\label{eq_appendix:TV}
	P_{\theta_{0}}^{(n)}\left[\Pi_{\boldsymbol{w}}\left(\theta\in\Theta:\,\,\|\theta-\theta_0\|\ge M_{n}\epsilon_{n}\mid\Xn\right)\mathbb{I}\left(\bm w_n\in R_n(c_0)\right)\right]
	\rightarrow0.
	\end{equation}
	Recall that Lemma \ref{lemma:unique_minimizer} says $\theta_0=\theta_{\bm w}^*$. Thereby, in Equation \eqref{eq_appendix:TV}, the usual center $\theta_{\bm w}^*$ to which the posterior $\Pi_{\bm w}\left(\cdot\mid X^{(n)}\right)$ concentrates is replaced by  $\theta_0$ (which does not depend on $\bm w$). 
	The second term $U_2$ in \eqref{eq_appendix:thm_together} can be trivially bounded using Equation \eqref{eq:weight_regularization}:
	\begin{equation}\label{eq_appendix:thm_bound2}
	P_{\boldsymbol{w}}\left[\mathbb{I}\left(\boldsymbol{w}\notin R_{n}\right)\right]=\mathbb{P}_{\boldsymbol{w}}\left(\boldsymbol{w}\notin R_{n}\right)\rightarrow0.
	\end{equation}
	Plugging \eqref{eq_appendix:TV} and \eqref{eq_appendix:thm_bound2} into \eqref{eq_appendix:thm_together}, we get the desired conclusion.

\end{proof}

\subsection{Proof of Theorem \ref{thm:mode}}\label{sec_appendix:proof_mode}
\begin{proof}
	We will prove the case where $\bm{w}_{n}=(w_{1},\cdots,w_{n})'\sim n\times Dir(c,\cdots,c)$.
	The other case is similar and easier.
	First, note that if we choose $z_{i}\stackrel{\text{iid}}{\sim}Gamma(c,1)$,
	we have
	\[
	\left(\frac{z_{1}}{\sum_{i=1}^{n}z_{i}},\cdots,\frac{z_{n}}{\sum_{i=1}^{n}z_{i}}\right)\sim Dir(c,\cdots,c).
	\]
	This implies
	\[
	\hat{\theta}_{\bm{w}}=\arg\min_{\theta}\left\{ \sum_{i=1}^{n}w_{i}\log p_{\theta}(X_{i})+\log\pi(\theta)\right\} 
	\]
	and 
	\begin{equation}\label{eq_appendix:thetahat_z}
	\hat{\theta}_{\bm{z}} =\arg\min_{\theta}\left\{ \sum_{i=1}^{n}z_{i}\log p_{\theta}(X_{i})+\frac{1}{n}\sum_{i=1}^{n}z_{i}\log\pi(\theta)\right\},\quad\text{where }\bm{z}=(z_{1},\cdots,z_{n})
	\end{equation}
	follows the same distribution. Thus, we focus on showing for any sequence $M_n\rightarrow\infty$,
	\[
	P_{{0}}^{(n)}\left[\P_{\bm{z}}\left(d(\hat{\theta}_{\bm{z}},\theta_{0})>M_{n}\epsilon_{n}\mid X^{(n)}\right)\right]\rightarrow0.
	\]
	This is equivalent to showing for any sequence $M_n\rightarrow\infty$,
	\[
	P_{{0}}^{(n)}\left[\P_{\bm{z}}\left(d(\hat{\theta}_{\bm{z}},\theta_{0})>2^{M_n}\epsilon_{n}\mid X^{(n)}\right)\right]\rightarrow0.
	\]
	Note that $\hat{\theta}_{\bm{z}}$ is a random variable where randomness comes from both $\bm z$ and $X^{(n)}$. Since $\bm z$ and $X^{(n)}$ are independent, denote the joint distribution of $X^{(n)}$ and $\bm z$ as $P_{0,\bm z}^{(n)}$, i.e.,
	\[
	\left(X^{(n)},\bm z\right)\sim P_{0,\bm z}^{(n)},\quad\text{where }dP_{0,\bm z}^{(n)}=dP_0^{(n)}\times dP_{\bm z}.
	\]
	And write
	\[
	\P^{(n)}_{{0},\bm{z}}(A)=\int_AdP_{0,\bm z}^{(n)}.
	\]
	Notice that from tower law, we have
	\[
	\P^{(n)}_{{0},\bm{z}}\left(d(\hat{\theta}_{\bm{z}},\theta_{0})>2^{M_n}\epsilon_{n}\right)=
	P_{{0}}^{(n)}\left[\P_{\bm{z}}\left(d(\hat{\theta}_{\bm{z}},\theta_{0})>2^{M_n}\epsilon_{n}\mid X^{(n)}\right)\right].
	\] 
	Define the shells \[S_{j}=\left\{ \theta\in\Theta:2^{j-1}\epsilon_{n}<d(\theta,\theta_{0})\le2^{j}\epsilon_{n}\right\},\quad \text{for } j=1,2,\dots\]
	Then we have, 
	\begin{align*}
	& \P^{(n)}_{0,\bm{z}}\left(d(\hat{\theta}_{\bm{z}},\theta_{0})>2^{M_n}\epsilon_{n}\right)\\
	= & \P^{(n)}_{0,\bm{z}}\left(d(\hat{\theta}_{\bm{z}},\theta_{0})>2^{M_n}\epsilon_{n},\sum_{i=1}^n z_{i}<2cn\right)+\P^{(n)}_{0,\bm{z}}\left(d(\hat{\theta}_{\bm{z}},\theta_{0})>2^{M_n}\epsilon_{n},\sum_{i=1}^n z_{i}\ge 2cn\right)\\
	\le & \sum_{j>M_n}\P^{(n)}_{0,\bm{z}}\left(2^{j-1}\epsilon_{n}<d(\hat{\theta}_{\bm{z}},\theta_{0})\le2^{j}\epsilon_{n},\sum z_{i}<2cn\right)+\P_{\bm{z}}\left(\sum_{i=1}^n z_{i}\ge 2cn\right)\\
	= & \sum_{j>M_n}\P^{(n)}_{0,\bm{z}}\left(\hat{\theta}_{\bm{z}}\in S_{j},\sum_{i=1}^n z_{i}<2cn\right)+\P_{\bm{z}}\left(\sum_{i=1}^n z_{i}\ge 2cn\right).
	\end{align*}
	Notice that by the weak law of large numbers, $\bar z_n=n^{-1}\sum_{i=1}^n z_i\xrightarrow{p}\E z_1=c$, and thus, 
	\[
	\P_{\bm{z}}\left(\sum_{i=1}^n z_{i}\ge 2cn\right)=\P_{\bm{z}}\left(\bar z_n\ge 2c\right)
	\le \P_{\bm{z}}\left(|\bar z_n-c|\ge c\right)\rightarrow0.
	\]
	Now, we will bound $\P^{(n)}_{0,\bm{z}}\left(\hat{\theta}_{\bm{z}}\in S_{j},\sum z_{i}<2cn\right)$
	for each $j>M_n$. Denote 
	\[
	\mathbb{M}_{n}(\theta)=\frac{1}{n}\sum_{i=1}^{n}z_{i}\log p_{\theta}(X_{i}),\quad \mathbb{M}(\theta)=P_{0}\log p_{\theta}(X_{i}).
	\]
	Recall  the definition of $\hat\theta_{\bm z}$ in Equation \eqref{eq_appendix:thetahat_z}. Thus, $\hat\theta_{\bm z}\in S_j$ implies 
	\[
	\sup_{\theta\in S_j}\left(\mathbb{M}_n(\theta)+\frac{1}{n^2}\sum_{i=1}^n z_i\log \pi(\theta)\right)\ge
	\mathbb{M}_n(\theta_0)+\frac{1}{n^2}\sum_{i=1}^n z_i\log \pi(\theta_0).
	\]
	Thus,
	\begin{align*}
	&\P^{(n)}_{0,\bm{z}}\left(\hat{\theta}_{\bm{z}}\in S_{j},\sum_{i=1}^n z_{i}<2cn\right)
	\\\le & \P^{(n)}_{0,\bm{z}}\left(\sup_{\theta\in S_{j}}\left(\mathbb{M}_{n}(\theta)-\mathbb{M}_{n}(\theta_{0})+\frac{1}{n^{2}}\sum_{i=1}^{n}z_{i}\log\frac{\pi(\theta)}{\pi(\theta_{0})}\right)\ge0,\sum_{i=1}^n z_{i}<2cn\right)\\
	\le & \P^{(n)}_{0,\bm{z}}\left(\sup_{\theta\in S_{j}}\left(\mathbb{M}_{n}(\theta)-\mathbb{M}_{n}(\theta_{0})+\frac{2c}{n}\log\frac{\pi(\theta)}{\pi(\theta_{0})}\right)\ge0\right).
	\end{align*}
	We have
	\begin{align}
	& \P^{(n)}_{0,\bm{z}}\left(\sup_{\theta\in S_{j}}\left(\mathbb{M}_{n}(\theta)-\mathbb{M}_{n}(\theta_{0})+\frac{2c}{n}\log\frac{\pi(\theta)}{\pi(\theta_{0})}\right)\ge0\right)\nonumber
	\\
	\le & \P^{(n)}_{0,\bm{z}}\left(\sup_{\theta\in S_{j}}\left(\mathbb{M}_{n}(\theta)-\mathbb{M}(\theta)-\mathbb{M}_{n}(\theta_{0})+\mathbb{M}(\theta_{0})+\frac{2c}{n}\log\frac{\pi(\theta)}{\pi(\theta_{0})}\right)+\sup_{\theta\in S_{j}}\left(-\mathbb{M}(\theta_{0})+\mathbb{M}(\theta)\right)\ge0\right)\nonumber
	\\
	= & \P^{(n)}_{0,\bm{z}}\left(\sup_{\theta\in S_{j}}\left(\mathbb{M}_{n}(\theta)-\mathbb{M}(\theta)-\mathbb{M}_{n}(\theta_{0})+\mathbb{M}(\theta_{0})+\frac{2c}{n}\log\frac{\pi(\theta)}{\pi(\theta_{0})}\right)\ge\inf_{\theta\in S_{j}}P_0\log[p_{\theta_0}/p_\theta]\right)\nonumber
	\\
	\stackrel{(a)}{\le} & \P^{(n)}_{0,\bm{z}}\left(\sup_{\theta\in S_{j}}\left(\left[\mathbb{M}_{n}(\theta)-\mathbb{M}(\theta)\right]-\left[\mathbb{M}_{n}(\theta_{0})-\mathbb{M}(\theta_{0})\right]\right)+\frac{2c}{n}\sup_{\theta\in S_{j}}\log\frac{\pi(\theta)}{\pi(\theta_{0})}\ge\inf_{\theta\in S_{j}}d^{2}(\theta_{0},\theta)\right)\nonumber
	\\
	\stackrel{(b)}{\le} & \P^{(n)}_{0,\bm{z}}\left(\sup_{\theta\in S_{j}}\left(\left[\mathbb{M}_{n}(\theta)-\mathbb{M}(\theta)\right]-\left[\mathbb{M}_{n}(\theta_{0})-\mathbb{M}(\theta_{0})\right]\right)\ge2^{2j-2}\epsilon_{n}^{2}-\frac{2c}{n}\sup_{\theta\in S_{j}}\log\frac{\pi(\theta)}{\pi(\theta_{0})}\right)\nonumber
	\\
	\stackrel{(c)}{\le} & \P^{(n)}_{0,\bm{z}}\left(\sup_{\theta\in S_{j}}\left(\left[\mathbb{M}_{n}(\theta)-\mathbb{M}(\theta)\right]-\left[\mathbb{M}_{n}(\theta_{0})-\mathbb{M}(\theta_{0})\right]\right)\ge2^{2j-3}\epsilon_{n}^{2}\right)\nonumber
	\\
	\stackrel{(d)}{\le} & \frac{P^{(n)}_{0,\bm{z}}\left[\sup_{\theta\in S_{j}}\left|\left[\mathbb{M}_{n}(\theta)-\mathbb{M}(\theta)\right]-\left[\mathbb{M}_{n}(\theta_{0})-\mathbb{M}(\theta_{0})\right]\right|\right]}{2^{2j-3}\epsilon_{n}^{2}},
	\label{eq_appendix:mode_markov}
	\end{align}
	where $(a)$ follows from Assumption (1), $(b)$ follows the definition
	of $S_{j}$, $(c)$ follows from the second part of \eqref{eq:mode_rate}, which says when $n$
	is sufficiently large, the following holds uniformly for all $j\ge 1$,
	\[
	2^{2j-2}\epsilon_{n}^{2}-\frac{2c}{n}\sup_{\theta\in S_{j}}\log\frac{\pi(\theta)}{\pi(\theta_{0})}>2^{2j-2}\epsilon_{n}^{2}-\frac{1}{2}2^{2j-2}\epsilon_{n}^{2}=2^{2j-3}\epsilon_{n}^{2},
	\]
	and $(d)$ follows from Markov's inequality.
	Now, our task is to bound the numerator in \eqref{eq_appendix:mode_markov}.
	Notice that 
	\begin{align*}
	& P^{(n)}_{0,\bm{z}}\left[\sup_{\theta\in S_{j}}\left|\left[\mathbb{M}_{n}(\theta)-\mathbb{M}(\theta)\right]-\left[\mathbb{M}_{n}(\theta_{0})-\mathbb{M}(\theta_{0})\right]\right|\right]\\
	= & P^{(n)}_{0,\bm{z}}\left[\sup_{\theta\in S_{j}}\left|\frac{1}{n}\sum_{i=1}^{n}z_{i}\log\frac{p_{\theta}(X_{i})}{p_{\theta_{0}}(X_{i})}-P_{0}\log\frac{p_{\theta}(X_{i})}{p_{\theta_{0}}(X_{i})}\right|\right]\\
	= & P^{(n)}_{0,\bm{z}}\left[\sup_{\theta\in S_{j}}\left|\frac{1}{n}\sum_{i=1}^{n}\left(z_{i}-1\right)\log\frac{p_{\theta}(X_{i})}{p_{\theta_{0}}(X_{i})}+\frac{1}{n}\sum_{i=1}^{n}\log\frac{p_{\theta}(X_{i})}{p_{\theta_{0}}(X_{i})}-P_{0}\log\frac{p_{\theta}(X_{i})}{p_{\theta_{0}}(X_{i})}\right|\right]\\
	\le & P^{(n)}_{0,\bm{z}}\left[\sup_{\theta\in S_{j}}\left|\frac{1}{n}\sum_{i=1}^{n}\left(z_{i}-1\right)\log\frac{p_{\theta}(X_{i})}{p_{\theta_{0}}(X_{i})}\right|\right]+P^{(n)}_{0}\left[\sup_{\theta\in S_{j}}\left|\frac{1}{n}\sum_{i=1}^{n}\log\frac{p_{\theta}(X_{i})}{p_{\theta_{0}}(X_{i})}-P_{0}\log\frac{p_{\theta}(X_{i})}{p_{\theta_{0}}(X_{i})}\right|\right].
	\end{align*}
	We will bound the two terms in the last line separately.
	
	First, from Lemma 2.9.1 in \cite{wellner2013weak} and Assumption (2),  
	we have
	\begin{align*}
	& P^{(n)}_{0,\bm{z}}\left[\sup_{\theta\in S_{j}}\left|\frac{1}{n}\sum_{i=1}^{n}\left(z_{i}-1\right)\log\frac{p_{\theta}(X_{i})}{p_{\theta_{0}}(X_{i})}\right|\right]\\
	\le & 2\sqrt{2n^{-1}}\|z_{1}-1\|_{2,1}\max_{k\in[n]}P^{(k)}_{0,\bm{z}}\left[\sup_{\theta\in S_{j}}\left|\frac{1}{\sqrt{k}}\sum_{i=1}^{k}\sigma_{i}\log\frac{p_{\theta}(X_{i})}{p_{\theta_{0}}(X_{i})}\right|\right]\\
	\le & C_{1}n^{-1/2}\max_{k\in[n]}\sqrt{k}\mathcal{R}_{k}\left(\mathcal{M}_{2^{j}\epsilon_{n}}\left(\theta_{0}\right)\right).
	\end{align*}
	By symmetrization inequality (Lemma 2.3.1 in \cite{wellner2013weak}),
	\[
	P^{(n)}_{0}\left[\sup_{\theta\in S_{j}}\left|\frac{1}{n}\sum_{i=1}^{n}\log\frac{p_{\theta}(X_{i})}{p_{\theta_{0}}(X_{i})}-P_{0}\log\frac{p_{\theta}(X_{i})}{p_{\theta_{0}}(X_{i})}\right|\right]\le2\mathcal{R}_{n}\left(\mathcal{M}_{2^{j}\epsilon_{n}}\left(\theta_{0}\right)\right).
	\]
	Thus, we have
	\begin{align*}
	& P^{(n)}_{0,\bm{z}}\left[\sup_{\theta\in S_{j}}\left|\left[\mathbb{M}_{n}(\theta)-\mathbb{M}(\theta)\right]-\left[\mathbb{M}_{n}(\theta_{0})-\mathbb{M}(\theta_{0})\right]\right|\right]\\
	\le & C_{1}\frac{\max_{k\in[n]}\sqrt{k}\mathcal{R}_{k}\left(\mathcal{M}_{2^{j}\epsilon_{n}}\left(\theta_{0}\right)\right)}{\sqrt{n}}+2\mathcal{R}_{n}\left(\mathcal{M}_{2^{j}\epsilon_{n}}\left(\theta_{0}\right)\right)\\
	\le & C_{2}n^{-1/2}\max_{k\in[n]}\phi_{k}(2^{j}\epsilon_{n}).
	\end{align*}
	Plugging into \eqref{eq_appendix:mode_markov}, we get 
	\[
	\P^{(n)}_{0,\bm{z}}\left(\sup_{\theta\in S_{j}}\left(\mathbb{M}_{n}(\theta)-\mathbb{M}_{n}(\theta_{0})+\frac{2c}{n}\log\frac{\pi(\theta)}{\pi(\theta_{0})}\right)\ge0\right)\le\frac{C_{2}n^{-1/2}\max_{k\in[n]}\phi_{k}(2^{j}\epsilon_{n})}{2^{2j-3}\epsilon_{n}^{2}}.
	\]
	By assumption, 
	$
	\phi_{k}(2^{j}\epsilon_{n})\le2^{\gamma j}\phi_{k}(\epsilon_{n})
	$
	and thus
	\[
	\P^{(n)}_{0,\bm{z}}\left(d(\hat{\theta}_{\bm{z}},\theta_{0})>2^{M}\epsilon_{n}\right)\le C_{4}\frac{\max_{k\in[n]}\phi_{k}(\epsilon_{n})}{8\sqrt{n}\epsilon_{n}^{2}}\sum_{j>M_n}2^{(\gamma-2)j}.
	\]
	As $M_n\rightarrow\infty$, we have
	$
	\sum_{j>M_n}2^{(\gamma-2)j}\rightarrow0.
	$
	And from assumption (4), we have
	\[
	\frac{\max_{k\in[n]}\phi_{k}(\epsilon_{n})}{\sqrt{n}\epsilon_{n}^{2}}\le C_{5},
	\]
	thus,
	\[
	C_{4}\frac{\max_{k\in[n]}\phi_{k}(\epsilon_{n})}{8\sqrt{n}\epsilon_{n}^{2}}\sum_{j>M_n}2^{(\gamma-2)j}\rightarrow0,
	\]
	which implies
	$
	\P^{(n)}_{0,\bm{z}}\left(d(\hat{\theta}_{\bm{z}},\theta_{0})>2^{M_n}\epsilon_{n}\right)\rightarrow0.
	$
\end{proof}

\section{Additional Experimental Results}
This section presents additional experimental results mentioned in the main text. We include investigations into  (a) the influence of network complexity, (b) the influence of regularization strength, and (c) additional results on Bayesian SVM and Bayesian LAD.

\subsection{Influence of Network Complexity}\label{sec_appendix:complexity}
We investigate the influence of complexity of the deep neural network on the approximating performance of the Deep Bootstrap sampler. We consider both the influence from the depth and the width of the architecture. Results are summarized in Table \ref{table:arch} where Table \ref{table:depth} investigates  the depth while Table \ref{table:width} investigates the width. We observe that in both tables,   the Deep Bootstrap sampler is not very sensitive to the complexity. As the depth (or width) further increases, the marginal gain from increasing the complexity becomes less and less important.

\begin{table}[H]
	\centering
	\small
	\begin{subtable}[t]{\textwidth}\centering
		\scalebox{0.7}{
			\begin{tabular}{l|l|l|l|l|l|l|l||l|l|l|l|l|l} 
				\hline\hline
				\bf \large Setting & &\multicolumn{6}{l||}{  \cellcolor[gray]{0.8}\large \bf $n=100,\,\,p=8$}  &\multicolumn{6}{l}{  \cellcolor[gray]{0.8}\large \bf $n=1\,000,\,\,p=50$}  
				\\
				\cline{2-14}
				& metric  & \multicolumn{2}{l|}{coverage} & \multicolumn{2}{l|}{length of 90\% CI} & \multicolumn{2}{l||}{bias}  
				& \multicolumn{2}{l|}{coverage} & \multicolumn{2}{l|}{length of 90\% CI} & \multicolumn{2}{l}{bias}  
				\\ \cline{2-14}
				&          width        & $+$     & $-$                     & $+$     & $-$                              & $+$     & $-$  & $+$     & $-$                     & $+$     & $-$                              & $+$     & $-$  
				\\ 
				\hline
				\multirow{2}{*}{DBS}   & 2                 & 0.90 &      0.88            &3.27  &    3.64                   & 0.80 &    0.92 
				& 0.88 &      0.85            &0.85  &    0.94                 & 0.23 &    0.25 
				\\
				& 3                   & 0.90 &      0.88     & 3.34 &          3.75         &  0.79&         0.92  
				& 0.88 &      0.86            &0.86  &    0.95                 & 0.23 &    0.25 
				\\ 
				& 4            &  0.90&    0.88      &  3.38&    3.78      & 0.79&    0.92   
				& 0.88 &      0.86            &0.87  &    0.95                 & 0.23 &    0.25  \\
				& 5           &  0.90&    0.88      &  3.41&    3.83             & 0.79&    0.92    
				& 0.88 &      0.87            &0.87  &    0.96                 & 0.23 &    0.25 
				\\ \hline
				WLB & -                  & 0.90 &      0.95      & 4.19&          4.71              &  0.79&         0.90    
				& 0.98 &      0.94      & 1.17&          1.24              &  0.23&         0.26  \\
				\hline\hline
			\end{tabular}
		}
		\caption{Fixing width (64 for $n=100,p=8$ and 128 for $n=1\,000,p=50$).}
		\label{table:depth}
	\end{subtable}
	
	\begin{subtable}[t]{\textwidth}\centering
		\scalebox{0.7}{
			\begin{tabular}{l|l|l|l|l|l|l|l||l|l|l|l|l|l} 
				\hline\hline
				\bf \large Setting & &\multicolumn{6}{l||}{  \cellcolor[gray]{0.8}\large \bf $n=100,\,\,p=8$}  &\multicolumn{6}{l}{  \cellcolor[gray]{0.8}\large \bf $n=1\,000,\,\,p=50$}  
				\\
				\cline{2-14}
				& metric  & \multicolumn{2}{l|}{coverage} & \multicolumn{2}{l|}{length of 90\% CI} & \multicolumn{2}{l||}{bias}  
				& \multicolumn{2}{l|}{coverage} & \multicolumn{2}{l|}{length of 90\% CI} & \multicolumn{2}{l}{bias}  
				\\ \cline{2-14}
				&          width        & $+$     & $-$                     & $+$     & $-$                              & $+$     & $-$  & $+$     & $-$                     & $+$     & $-$                              & $+$     & $-$  
				\\ 
				\hline
				\multirow{2}{*}{DBS}   & 16                 & 0.90 &      0.88            &3.00  &    3.37                   & 0.79 &    0.92 
				& 0.85 &      0.82            &0.80  &    0.88                   & 0.23 &    0.25 
				\\
				& 32                   & 0.90 &      0.88     & 3.17 &          3.53         &  0.80&         0.93  
				& 0.85 &      0.83            &0.81  &    0.89                   & 0.23 &    0.25 
				\\ 
				& 64            &  0.90&    0.88      &  3.29&    3.64      & 0.80&    0.92   
				& 0.88 &      0.84            &0.84  &    0.92                   & 0.23 &    0.25  \\
				& 128           &  0.90&    0.88      &  3.31&    3.72             & 0.79&    0.92   
				& 0.88 &      0.86            &0.86  &    0.95                   & 0.23 &    0.25  
				\\ \hline
				WLB & -                  & 0.90 &      0.95      & 4.19&          4.71              &  0.79&         0.90  
				& 0.98 &      0.94      & 1.17&          1.24              &  0.23&         0.26    \\
				\hline\hline
			\end{tabular}
		}
		\caption{Fixing depth (2 for $n=100,p=8$ and 3 for $n=1\,000,p=50$).}
		\label{table:width}
	\end{subtable}
	\caption{\small Sensitivity analysis for deep Bootstrap sampler in Bayesian LAD regression example. We consider Model 2 with different $(n,p)$'s.  `DBS' stands for deep Bootstrap sampler. `WLB' standards for samples from the true bootstrap distribution. Coverage stands for the empirical coverage of 90\% confidence itervals. `Bias'   refers to the $l_1$ distance between estimated posterior means and the truth. We denote with $+$ an average over active coordinates, and with $-$ an average over inactive coordinates. Results are averaged over 10 independently generated data sets.}
	\label{table:arch}
\end{table}

\subsection{Influence of Regularization Strength}
\label{sec_appendix:regularization}
In examples in the main text, we choose the regularization parameter ($\alpha$ for Bayesian NPL and $\lambda$ for Gibbs posterior) either subjectively or by consulting classic model selection criteria. One intriguing question is whether the regularization strength has an impact on approximating performance of the deep Bootstrap sampler, and if there is, how. This subsection investigates this question.

\paragraph{Influence of $\lambda$ in Gibbs posterior.}
Let us consider the classic LASSO regression example. Given data $\{(y_i,\bm x_i)\in\R\bigotimes\mathbb{R}^p\}_{i=1}^n$ where $\x_i$ denotes the covariates of an observation with response $y_i$ and assume a linear model, LASSO regression  solves the penalized regression problem 
\[
\sum_{i=1}^nl(\bm\theta;y_i,\x_i)+\lambda\sum_{j=1}^p|\theta_j|, 
\]
which is equivalent to a the MAP of a Gibbs posterior with loss
\[
l(\bm\theta;y_i,\x_i)=(y_i-\bm\theta^\top\x_i),
\]
and prior
\[
\pi(\bm\theta)=\prod_{j=1}^p\left[\frac{\lambda}{2}e^{-\lambda|\theta_j|}\right].
\]
Here we assume the covariates and response are both centered so intercept is not included. 
One can use BIC strategy similar to that in Section \ref{sec:LAD} to choose the regularization strength $\lambda$. Yet here we will investigate different choices of $\lambda$'s and its effect on the deep Bootstrap sampler.

We simulate data following the settings in \cite{nie2020bayesian}. We generate $n$ i.i.d. pairs $(\x_i,y_i)$ from
\[
y_i=\x_i^\top\bm\theta+\epsilon_i,\quad\x_i\stackrel{\text{iid}}{\sim}N(\bm 0_p,\Sigma),\quad \epsilon_i\stackrel{\text{iid}}{\sim}N(0,1).
\]
with $n=1\,000,p=50$, and $\rho$ a $p\times p$ matrix with all 1's on the diagonal and $\rho$'s off-diagonally. We consider $\rho=0.6$.

The Bootstrap truth is computed using the function \texttt{linear\_model.Lasso} in Python library \texttt{sklearn} \cite{scikit-learn}. 

\begin{figure}
	\centering
	\includegraphics[width=\textwidth,height=.4\textheight]{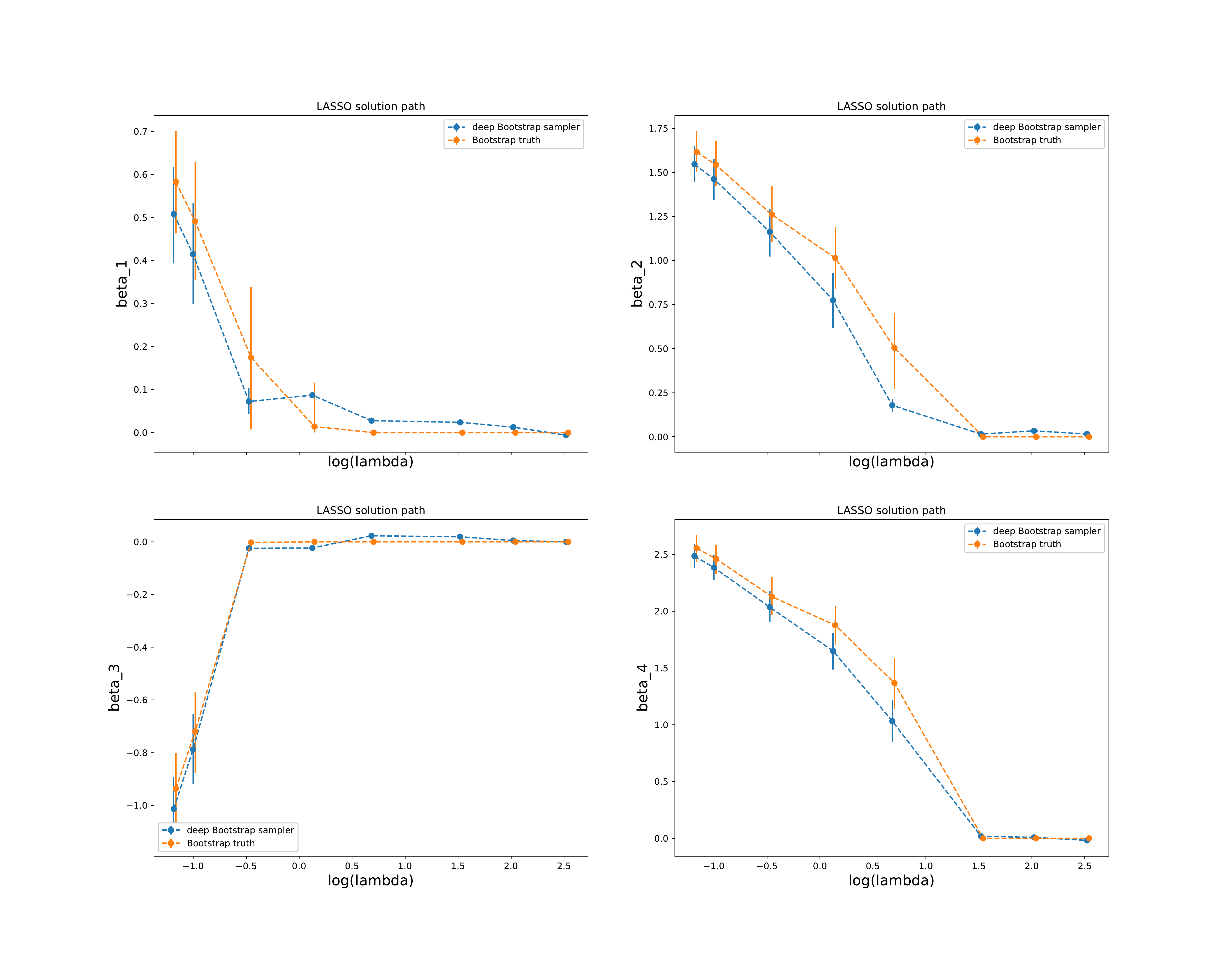}
	\caption{\small LASSO solution path for the 4 active coordinates from deep Bootstrap sampler (blue line) and Bootstrap truth (orange line). We set $n=1\,000,p=50,\rho=0.6$. The active coordinates are $\boldmath{\theta}_{active}=(1,2,-2,3)'$. 
		Each subfigure corresponds to one active coordinate.}
	\label{fig_appendix:LASSO_path}
\end{figure}

We plot the 95\% confidence interval from both the Bootstrap target and the deep Bootstrap sampler for different value of $\lambda$'s, i.e., the so-called `solution path'. The plot is shown in Figure  \ref{fig_appendix:LASSO_path}. 
We observe that in both correlation structures, as regularization parameter $\lambda$ grows, the degree of underestimation for deep Bootstrap sampler also grows.

\begin{figure}
	\centering
	\includegraphics[width=\textwidth,height=.4\textheight]{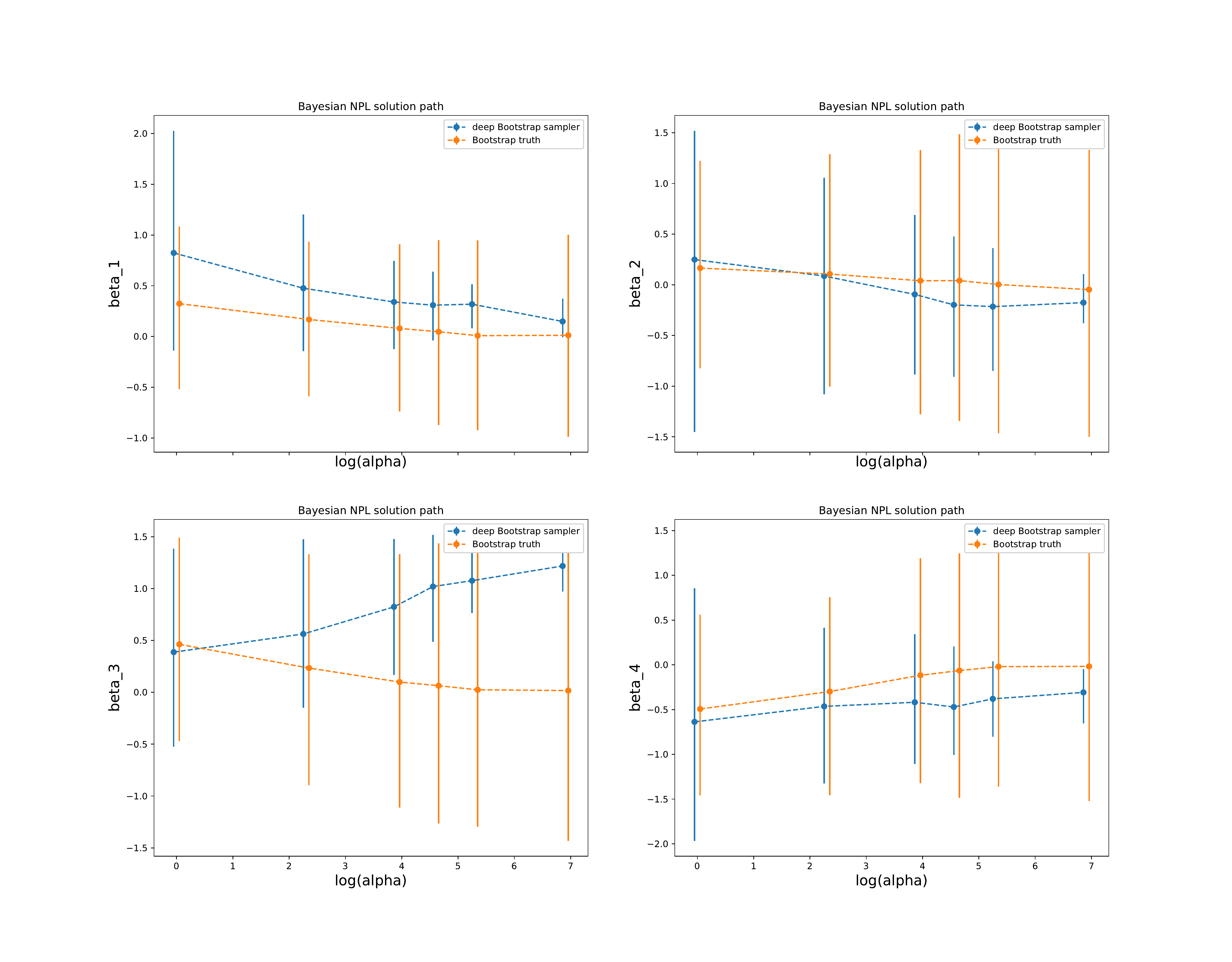}
	\caption{\small Solution path for the Bayesian support vector machine example. We set $n=50,p=10,\rho=0.6$. }
	\label{fig_appendix:NPL_path}
\end{figure}

\paragraph{Influence of $\alpha$ in Bayesian NPL.}
We consider also the Bayesian support vector machine example introduced in Section \ref{sec:svm}, with $\rho=0.6$. Note that in Bayesian NPL, the effect of $\alpha$ is analogous to $\lambda$ in Gibbs posterior. So we make a similar solution path for varying $\alpha$'s in Figure \ref{fig_appendix:NPL_path}. Again, we observe that as $\alpha$ increases, in general the degree of underestimation also increases, which is consistent to our observations in the previous example.

\paragraph{Conclusion.} We conclude that regularization strength has an impact on the performance of deep Bootstrap samplers. In general, the larger the regularization strength is, the more severe the variance underestimation problem is.

\subsection{Additional Results on Bayesian Support Vector Machine}
\label{sec_appendix:svm}

\begin{figure}[H]
	\centering
	\includegraphics[width=.45\linewidth, height=.15\textheight]{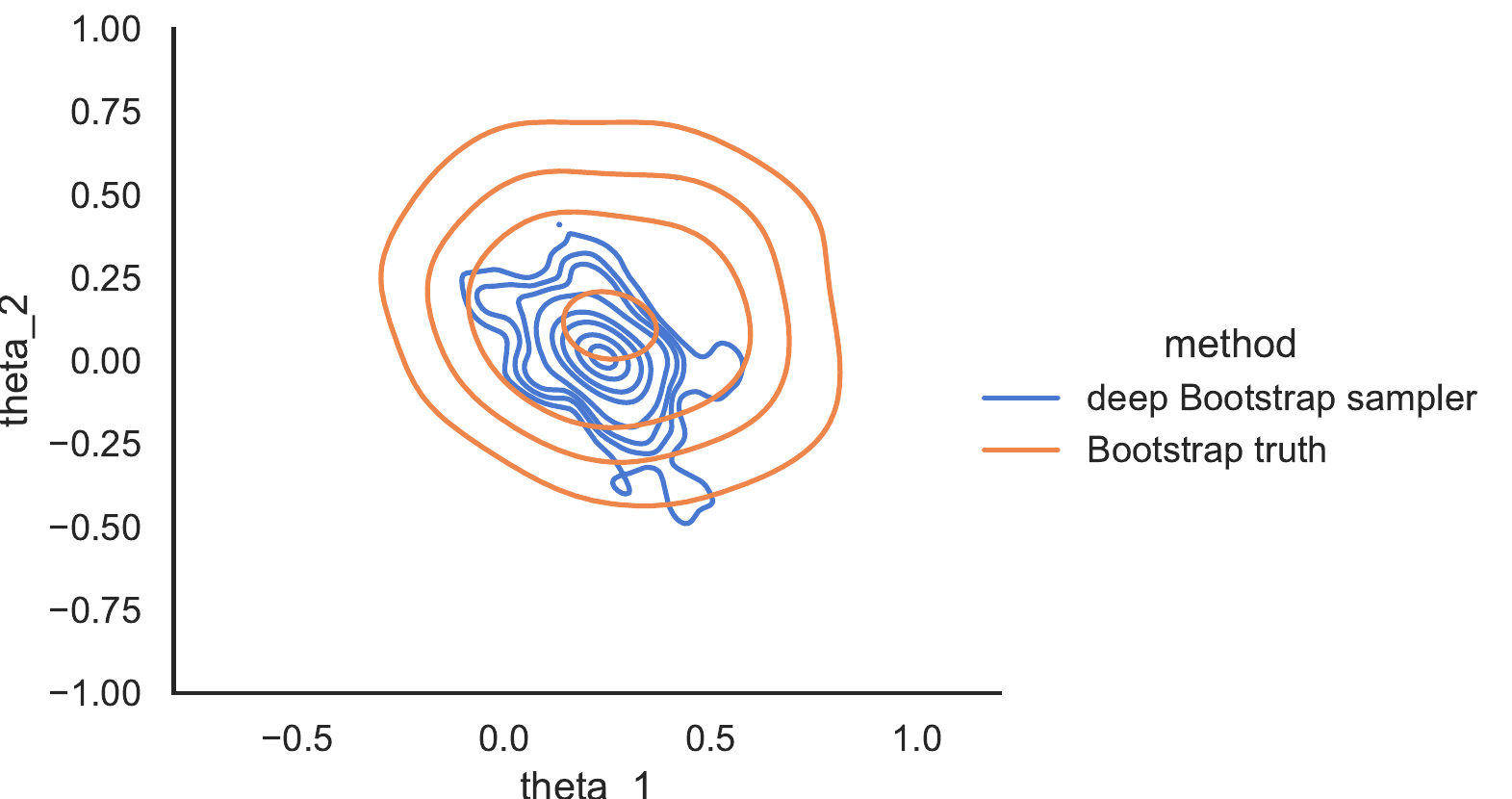}
	\includegraphics[width=.45\linewidth, height=.15\textheight]{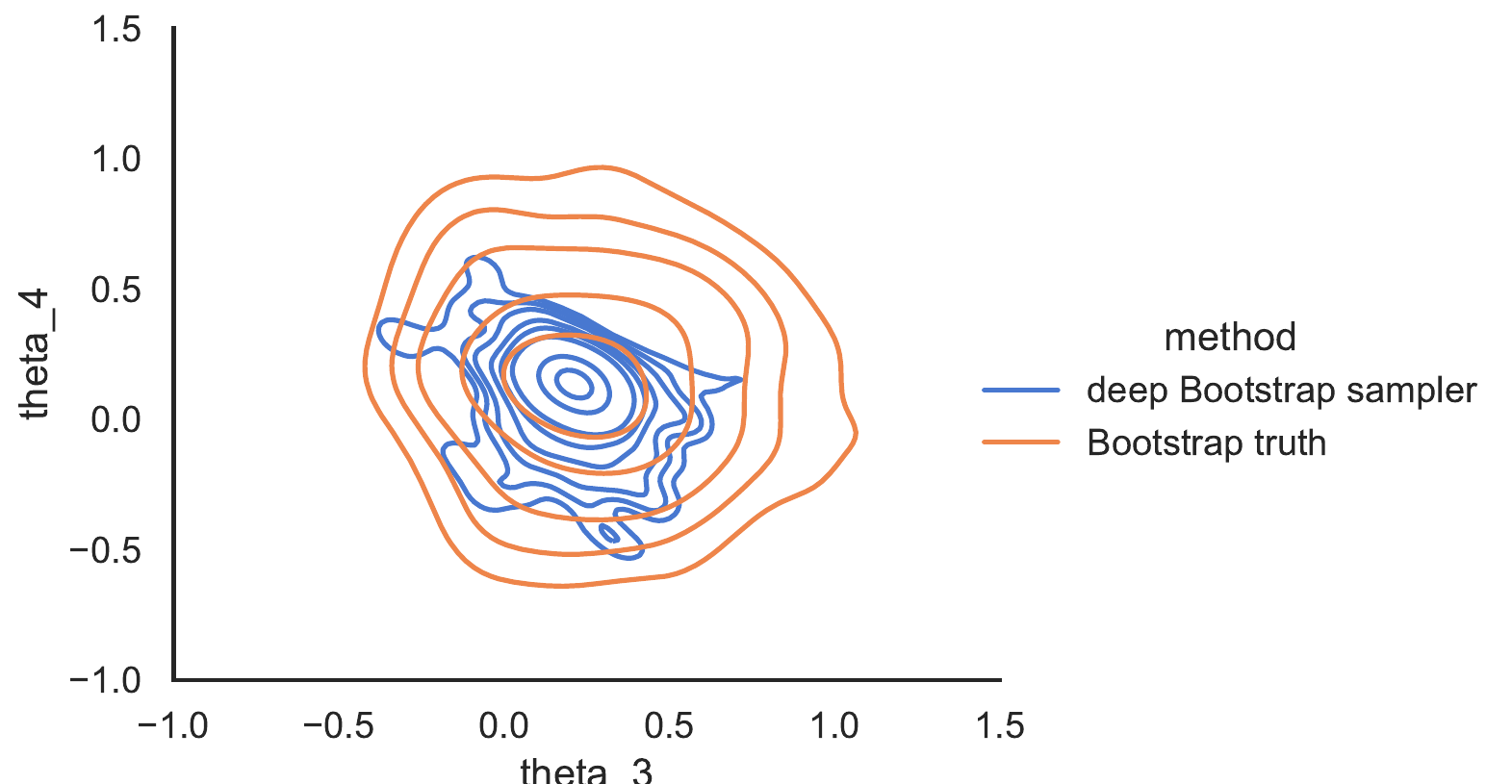}
	\caption{\small Two-dimensional posterior density plot for $\theta_i$'s from deep Bootstrap sampler and the truth for Bayesian support vector machine example. We set $n=50,p=10,\rho=0,\alpha=1.0$. 
	}
	\label{fig:NPL_2d_density_ind}
\end{figure}

We present results on the Bayesian support vector machine for the independent case ($\rho=0$). The two-dimensional density plots are shown in Figure \ref{fig:NPL_2d_density_ind}. We observe that the posterior mean is captured well, while variance is more underestimated compared to correlated case. 
It is because with independent covariates, this is a simple problem, 
{where similar and small values of $(\beta,\bm\theta)$ will always yield a small loss, regardless of the input weight $\bm w$. Consequently, variation in the input weight $\bm w$'s only turned into small variation over output $\hat G(\bm w)$.} 
We note that this issue does not affect the predictive performance, which is shown in Table \ref{table:svm_ind}.  

\begin{table}[H]
	\centering
	\small
	\scalebox{0.7}{
		\begin{tabular}{l|l|l|l|l|l|l|l|l} 
			\hline\hline
			\bf \large Setting&&\multicolumn{7}{|l}{  \cellcolor[gray]{0.8}\large \bf Equi-correlated $\rho=0.6$}    \\
			\cline{1-9}
			& metric   
			& accuracy & precision & recall & F1 & ROC & PR & time\\ 
			\hline
			\multirow{2}{*}{$p=10,n=50$}   & DBS                     
			& 0.99  &       1.00          & 0.99  &        0.99              &1.00  &   1.00 & 54.68+0.65\\
			& WLB               
			& 1.00  &   1.00     & 1.00 &    1.00                   &  1.00&      1.00     &176.20\\ 
			\hline
			\multirow{2}{*}{$p=50,n=500$}  & DBS       
			& 1.00 & 1.00& 1.00 &1.00 &1.00 &1.00 &100.92+0.65      \\
			& WLB      
			& 1.00 & 1.00& 1.00 &1.00 &1.00 &1.00    &199.98          \\ 
			\hline
			\multirow{2}{*}{$p=100,n=1000$}  & DBS                
			& 1.00 & 1.00& 1.00 &1.00 &1.00 &1.00  &256.87+0.87\\
			& WLB       
			& 1.00 & 1.00& 1.00 &1.00 &1.00 &1.00 &548.26\\ 
			\hline
			\multirow{2}{*}{$p=200,n=2000$} & DBS        
			& 1.00 & 1.00& 1.00 &1.00 &1.00 &1.00  & 378.68+0.80\\
			& WLB            
			& 1.00 & 1.00& 1.00 &1.00 &1.00 &1.00 &1130.23 \\
			\hline
			\multirow{2}{*}{$p=500,n=5000$} & DBS      
			& 1.00 & 1.00& 1.00 &1.00 &1.00 &1.00   & 559.22+0.64\\
			& WLB            
			& 1.00 & 1.00& 1.00 &1.00 &1.00 &1.00 & 6256.95 \\
			\hline\hline
		\end{tabular}
	}
	\caption{\small Evaluation of approximation properties based on 10 independent runs for Bayesian support vector machine example. DBS standards for `deep Bootstrap sampler'. WLB standards for samples from the true bootstrap distribution. `Bias'   refers to the $l_1$ distance of estimated posterior means. `ROC' refers to the area under the receiver operating characteristic curve;  `PR' refers to the area under the precision-recall curve. The last column in each setting represents the time (in seconds) to generate 10\,000 sample points. For deep Bootstrap sampler, time reported is in the form of training time $+$ sampling time.}
	\label{table:svm_ind}
\end{table}

{
	To confirm our hypothesis on the reason of variance underestimation in this example, we conduct the same experiments but on a more challenging setting where
	\begin{equation}\label{eq_appendix:svm_generation_difficult}
	\P\left(y_i=1\right)=\P\left(y_i=-1\right)=1/2, \quad \bm x_i\mid y_i\sim N(0.3y_i\bm 1_p,\Sigma),
	\end{equation}
	where $\Sigma$ is a $p\times p$ matrix with all $1$'s on the diagonal and $0$'s off-diagonally. We note that the signal strength in this model is smaller than that in the original setting \eqref{eq:svm_generation}, and thus this problem is more difficult in terms of prediction. We set the same prior strength $\alpha=1$ as the original example. 
	The posterior density plot for $\theta_i$'s from this model are shown in Figure \ref{fig:NPL_2d_density_ind_difficult}. Comparing Figure \ref{fig:NPL_2d_density_ind_difficult} with Figure \ref{fig:NPL_2d_density_ind}, we observe that in the easier setting, the variance underestimation issue is more severe, which confirms our previous hypothesis.
	
	\begin{figure}[H]
		\centering
		\includegraphics[width=.45\linewidth, height=.15\textheight]{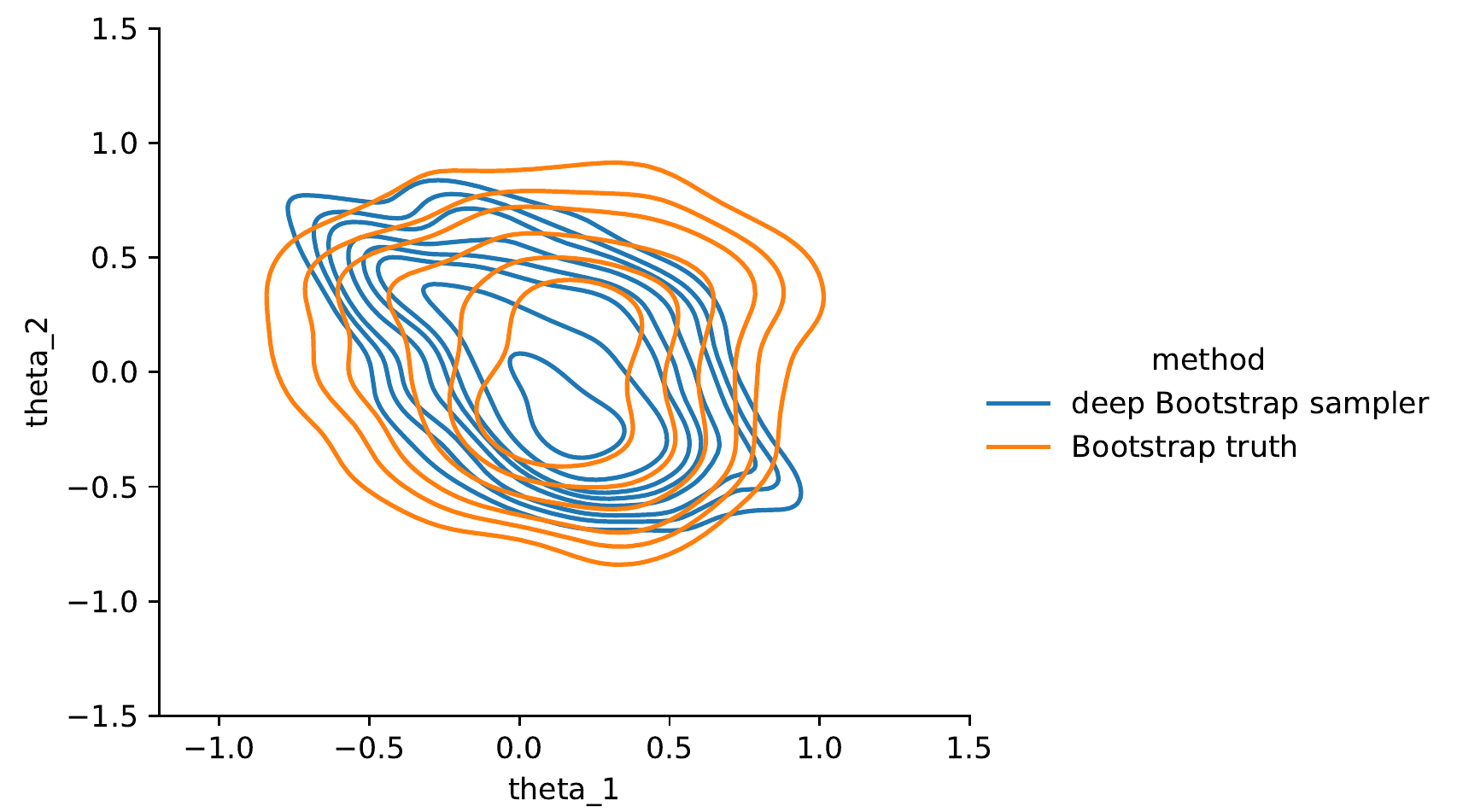}
		\includegraphics[width=.45\linewidth, height=.15\textheight]{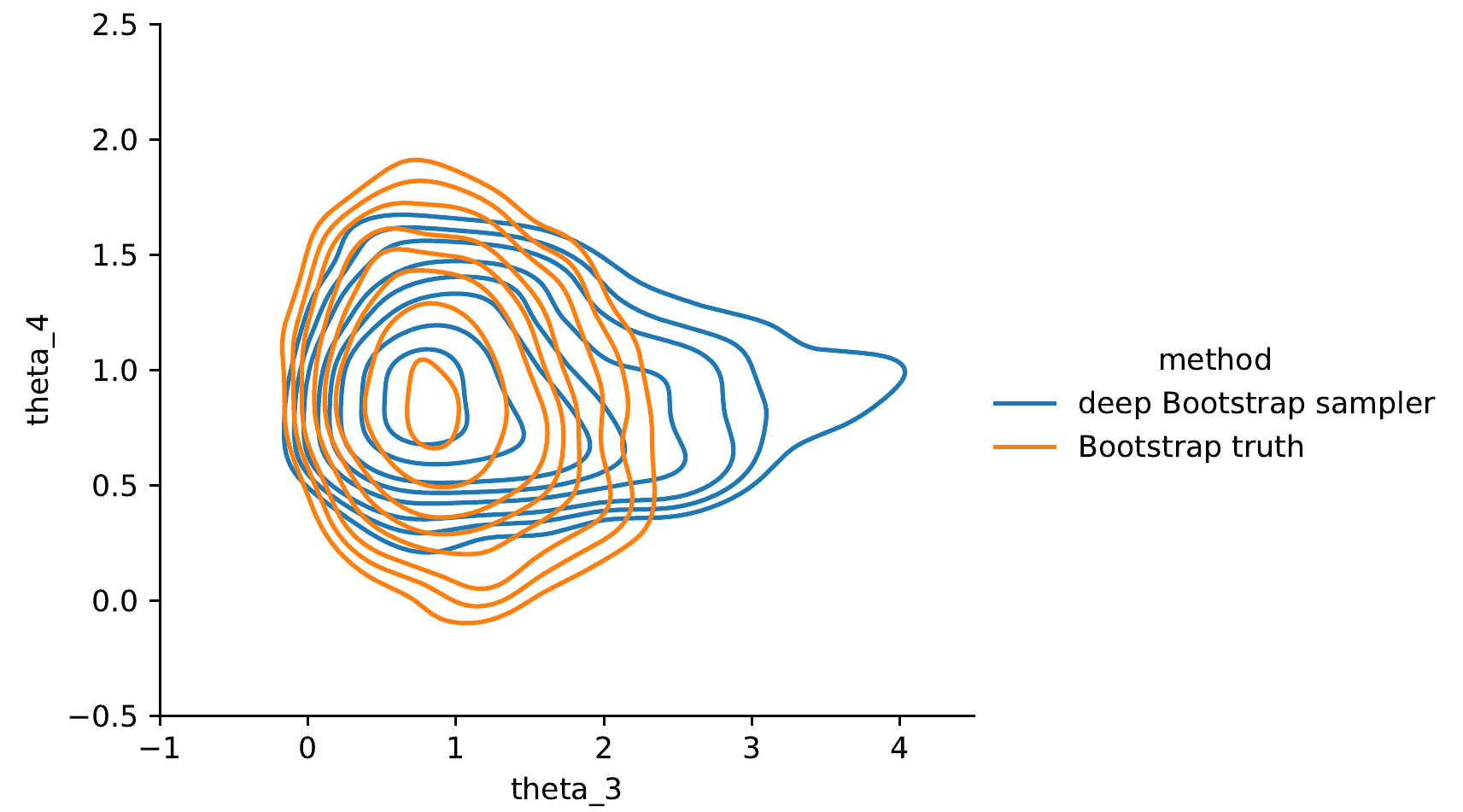}
		\caption{\small Two-dimensional posterior density plot for $\theta_i$'s from deep Bootstrap sampler and the truth for Bayesian support vector machine example. We set $n=50,p=10,\rho=0,\alpha=1.0$, and the data are generated using a more challenging setting \eqref{eq_appendix:svm_generation_difficult}.
		}
		\label{fig:NPL_2d_density_ind_difficult}
	\end{figure}
	
}

\subsection{Additional Results on Bayesian LAD Regression}\label{sec_appendix:LAD}
\begin{figure}[H]
	\centering
	\includegraphics[width=.45\linewidth, height=.15\textheight]{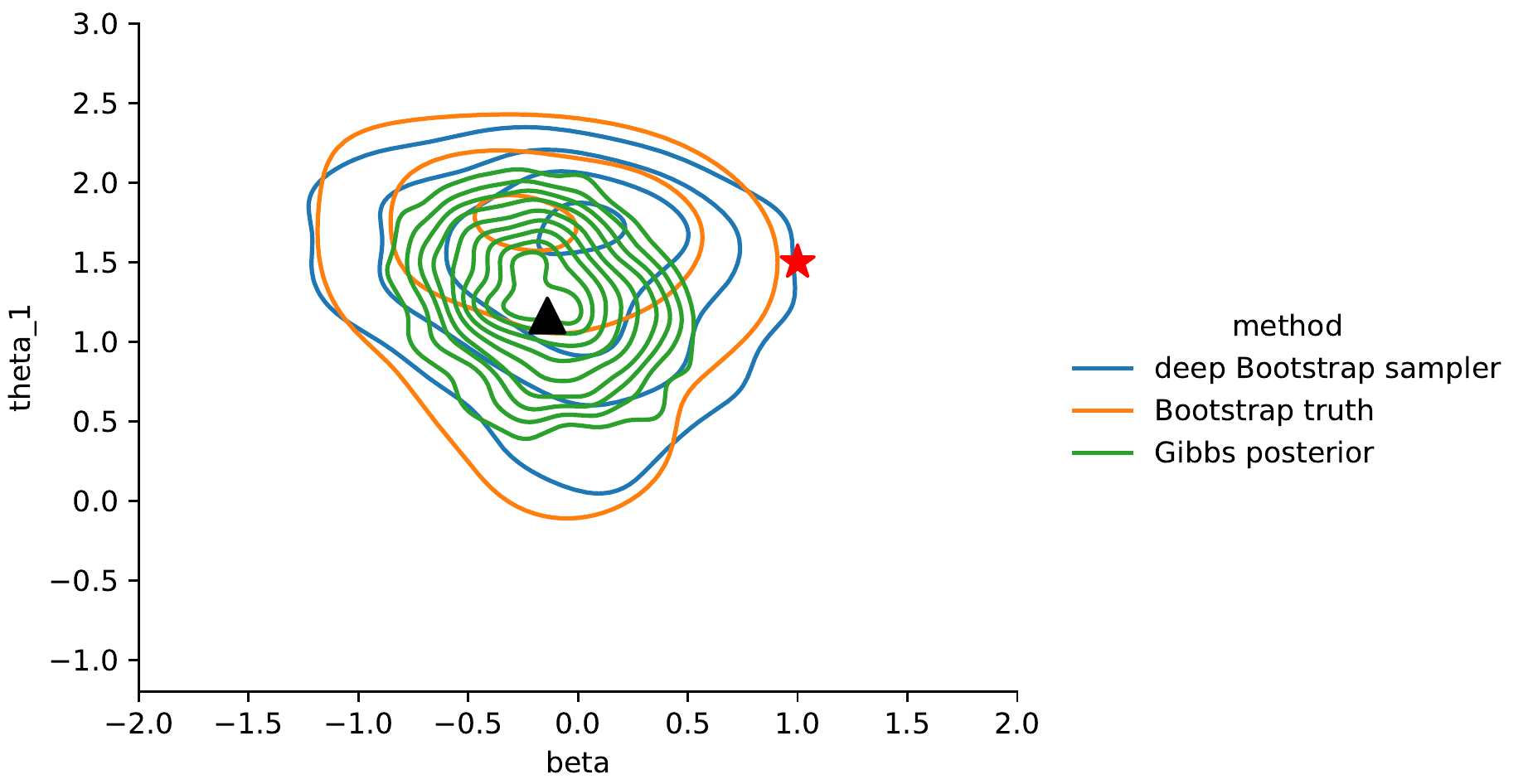}
	\includegraphics[width=.45\linewidth, height=.15\textheight]{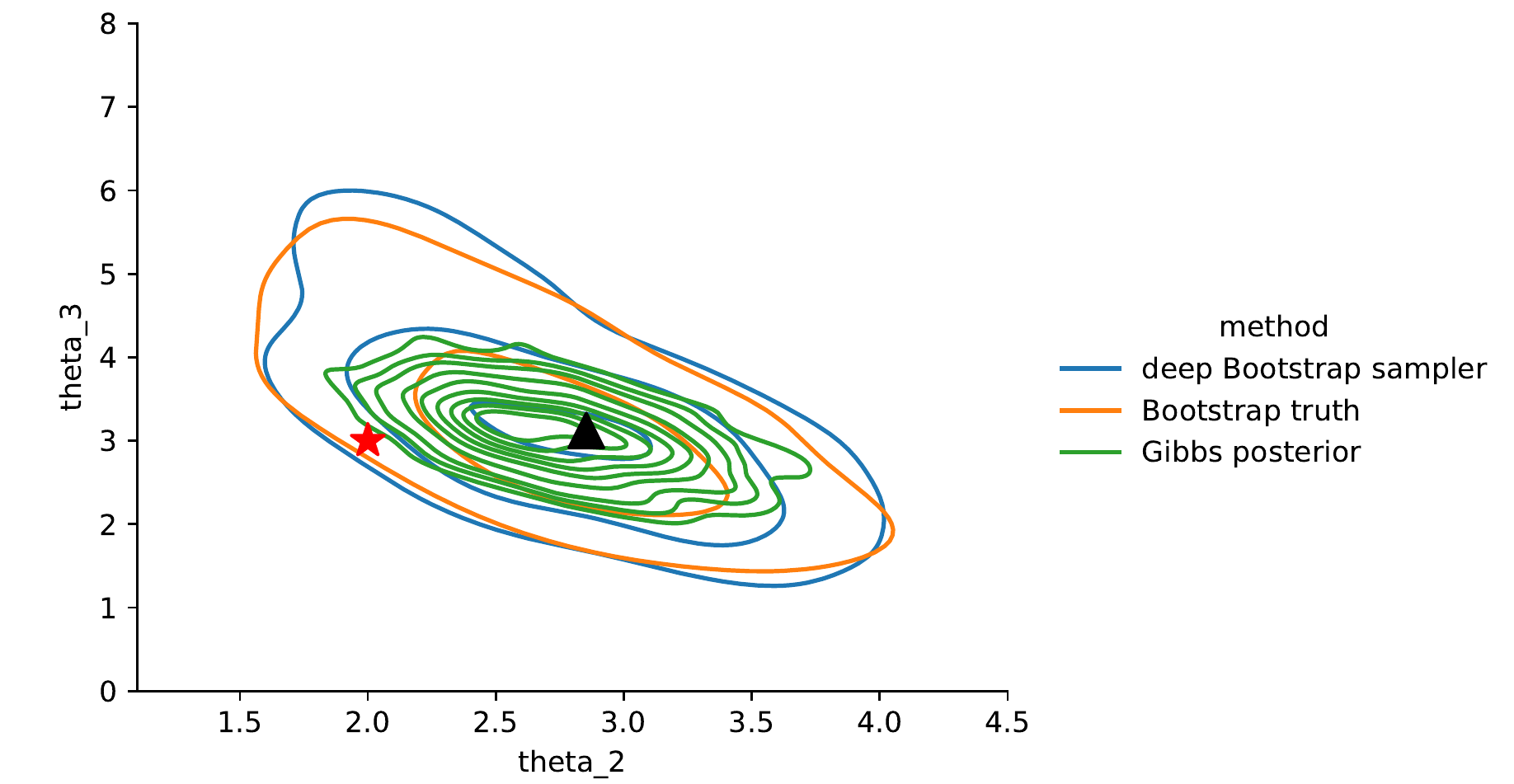}
	\caption{\small Two-dimensional posterior density plot for $\theta_i$'s from Model 2 for Bayesian LAD example. We set $n=100,p=8$. The location of true parameters are marked in red star. The location of the minimizer of Equation \eqref {eq:penalized_LAD} are marked in black triangle.}
	\label{fig:LAD_2d_density_sensible}
\end{figure}

Here we present additional results for Bayesian LAD regression, Model 2. The two-dimensional density plot is shown in Figure \ref{fig:LAD_2d_density_sensible}. Again we observe the same location for all methods, yet smaller variance for Gibbs posterior as compared to Bootstrap samples. Posterior from deep Bootstrap sampler is very close to the Bootstrap truth.

\end{document}